\documentclass[fleqn,usenatbib,aastex,preprintnumbers]{mnras}

\usepackage{newtxtext,newtxmath}
\usepackage[T1]{fontenc}
\usepackage{amsmath}
\usepackage{booktabs}

\DeclareRobustCommand{\VAN}[3]{#2}
\let\VANthebibliography\thebibliography
\def\thebibliography{\DeclareRobustCommand{\VAN}[3]{##3}\VANthebibliography}

 \usepackage{comment}
 \usepackage{titlesec}
\usepackage{url}
\titleformat{\paragraph}
{\normalfont\normalsize\bfseries}{\theparagraph}{1em}{}
\titlespacing*{\paragraph}
{0pt}{3.25ex plus 1ex minus .2ex}{1.5ex plus .2ex}

\usepackage{graphicx}	
\usepackage{amsmath}

\usepackage[normalem]{ulem}
\newcounter{tr}
\ifnum \value{tr}>5

\newcommand{\deletedD}[1]{{\color{purple} Damien - Deleted: } \sout{#1}}

\newcommand{\authorcommentD}[1]{{\color{purple} Damien - Comment :} {\color{cyan} #1}}

\else

\newcommand{\deletedD}[1]{}

\newcommand{\authorcommentD}[1]{}

\fi

\newenvironment{system}%
{\left\lbrace\begin{array}{@{}l@{}}}%
{\end{array}\right.}
\usepackage{tikz,xcolor,hyperref}

\definecolor{lime}{HTML}{A6CE39}
\DeclareRobustCommand{\orcidicon}{\hspace{-1mm}
	\begin{tikzpicture}
	\draw[lime, fill=lime] (0,0) 
	circle [radius=0.16] 
	node[white] {{\fontfamily{qag}\selectfont \tiny \,ID}};
	\draw[white, fill=white] (-0.0525,0.095) 
	circle [radius=0.007];
	\end{tikzpicture}
	\hspace{-3mm}
}

\foreach \x in {A, ..., Z}{\expandafter\xdef\csname orcid\x\endcsname{\noexpand\href{https://orcid.org/\csname orcidauthor\x\endcsname}
			{\noexpand\orcidicon}}
}

\title[Probing gamma-ray bursts observed at very high energies]{Probing gamma-ray bursts observed at very high energies through their afterglow}

\author[Guarini et al.]{
Ersilia Guarini\orcidA{}$^{1}$\thanks{E-mail:~ersilia.guarini@nbi.ku.dk},
Irene Tamborra\orcidB{}$^{1}$\thanks{E-mail:~tamborra@nbi.ku.dk},
Damien B{\'e}gu{\'e}\orcidC{}$^{2}$\thanks{E-mail:~begueda@biu.ac.il},
and Annika Rudolph\orcidD{}$^{1}$\thanks{E-mail:~annika.lena.rudolph@nbi.ku.dk} \\ 
\\
$^{1}$Niels Bohr International Academy \& DARK, Niels Bohr Institute, University of Copenhagen, Blegdamsvej 17, 2100, Copenhagen, Denmark \\
$^{2}$ Department of Physics, Bar Ilan University, 52900, Ramat-Gan, Israel
}

\pubyear{2023}

\begin{document}
\label{firstpage}
\pagerange{\pageref{firstpage}--\pageref{lastpage}}
\maketitle

\begin{abstract}
A growing number of gamma-ray burst (GRB) afterglows is observed at very-high energies (VHE, {$\gtrsim 100$~GeV}).
Yet, our understanding of the mechanism powering the VHE emission remains baffling. 
We make use of multi-wavelength observations of the afterglow of GRB 180720B, GRB 190114C, and GRB 221009A to investigate whether the bursts exhibiting VHE emission  share common features. We assume the standard afterglow model and microphysical parameters consistent with a synchrotron self-Compton (SSC) scenario for the VHE radiation. By requiring that the blastwave should be transparent to $\gamma$--$\gamma$ pair production at the time of observation of the VHE photons and relying on typical prompt emission efficiencies and data in the radio, optical and X-ray bands, we infer for those bursts that the initial energy of the blastwave is  $\tilde{E}_{k, \rm{iso}} \gtrsim  \mathcal{O}(10^{54})$~erg and the circumburst density is $n_0  \lesssim \mathcal{O}(10^{-1})$~cm$^{-3}$ for a constant  circumburst profile  [or $A_\star \lesssim \mathcal{O}(10^{-1})$~cm$^{-1}$ for a wind scenario]. Our findings thus suggest that these VHE bursts might be hosted in low-density environments, if the SSC radiation is responsible for the VHE emission. While these trends are based on a small number of bursts,  the Cherenkov Telescope Array has the potential to  provide crucial insight in this context by detecting a larger sample of VHE GRBs. In addition, due to the very poor statistics, the non-observation of high-energy neutrinos  cannot  constrain the  properties of these bursts efficiently, unless additional VHE GRBs should be detected  at distances closer than $15$~Mpc when IceCube-Gen2 radio will be operational. 
\end{abstract}

\begin{keywords}
gamma-ray bursts -- acceleration of particles -- ISM: jets and outflows 
\end{keywords}

\section{Introduction}
Gamma-ray bursts (GRBs) are among the most powerful explosions in our Universe~\citep{Klebesadel:1973iq, Kumar:2014upa, Piran:2004ba}. They exhibit a non-thermal spectrum with typical peak energies in the keV--MeV range~\citep{Poolakkil:2021jpc, vonKienlin:2020xvz, Ford:1994uy}.  
We focus on long-duration GRBs, which release an isotropic energy in gamma-rays of about $10^{49}$--$10^{55}$~ergs within a few $\mathcal{O}(10)$~s~\citep{Kumar:2014upa, Atteia:2017dcj}. The pulse of gamma-rays released during the prompt phase is followed by a delayed, long-lasting emission: the afterglow. Being detected from the radio to the X-ray bands, the afterglow makes GRBs electromagnetically accessible across all wavebands~\citep[e.g.,][]{Meszaros:1996sv}. 

Afterglow detections at high energy (HE, $\gtrsim 1$~GeV) have been reported for more than a decade, e.g.~by the Large Area Telescope (LAT) onboard of the \textit{Fermi} satellite~\citep{Ajello:2019zki}. 
In the past few years, such observations have been complemented by the detection of very high energy (VHE, {$\gtrsim 100$~GeV}) emission from an increasing number of GRBs, with  
photons with $\mathcal{O}(\rm{TeV})$ energy being detected several hours after the burst trigger~\citep{Abdalla:2019dlr, HESS:2021dbz, MAGIC:2019irs, MAGIC:2019lau, Suda:2021dox,MAGIC:2021jcq}. 
Among these puzzling bursts, the recently discovered GRB 221009A represents an extraordinary event, being located close-by ($z\simeq 0.15$),  very bright in gamma-rays ($\tilde{E}_{\gamma, \rm{iso}} \gtrsim 3 \times 10^{54}$~ergs)~\footnote{We use  three different reference frames throughout this paper: the observer frame, the central engine frame, and the blastwave comoving frame. In each of these frames, quantities are denoted with $X, \tilde{X}, X^{\prime}$, respectively.}, and detected with photons up to $\mathcal{O}(10)$~TeV by the Large High Altitude Air Shower Observatory (LHAASO)~\citep{2022GCN.32677....1H}.

The VHE emission associated with the GRB afterglow was theoretically predicted~\citep{Meszaros:1993ft, Dermer:1999eh,Meszaros:2001vi, Piran:2004ba, Kumar:2014upa}, and then observed thanks to ground-based Cherenkov telescopes, such as the High Energy Stereoscopic System (H.E.S.S.) and the Major Atmospheric Gamma Imaging Cherenkov (MAGIC).
The detection rate of GRB photons with energies $\gtrsim \mathcal{O}(\rm{TeV})$ is expected to further improve with LHAASO~\citep{2022GCN.32677....1H} and the upcoming Cherenkov Telescope Array (CTA)~\citep{Knodlseder:2020onx}; hence, it is timely to investigate under which conditions VHE emission should be expected. 

Up to HE, the multi-wavelength emission of the GRB afterglow is broadly considered to be generated by the synchrotron radiation produced by the electrons accelerated at the external shock  as  the latter expands in the circumburst medium (CBM)~\citep{Meszaros:1996sv, Waxman:1997ga, Waxman:1997if, 1997ApJ...490..772K, Sari:1997qe}. Yet, this standard afterglow picture cannot accommodate the production of TeV photons, unless electrons are accelerated above the synchrotron cut-off energy---see, e.g.,~\cite{HESS:2021dbz}.

A possibility proposed to explain the VHE emission is the synchrotron self-Compton (SSC) scenario, according to which synchrotron photons inverse-Compton scatter the electrons that produced them~\citep{Ghisellini:1998jy, Chiang:1998wh, Dermer:2000yd, Sari:2000zp,2009ApJ...703..675N, Liu:2013foa, Derishev:2021ivd, Fraija:2019rag, Asano:2020grw, Khangulyan:2023akp}. Alternatively, the acceleration of baryons together with electrons at the external shock can be considered. In this case, the mechanism responsible for the VHE emission may be proton synchrotron radiation or the decay of secondaries produced in photo-pion and photo-pair processes~\citep[e.g.,][]{Bottcher:1998qi, Asano:2008tc, Razzaque:2009rt, Gagliardini:2022rrq, Isravel:2022glo}. Photohadronic processes have also been invoked for modeling the VHE emission~\citep[e.g.,][]{Sahu:2020dsg, Sahu:2022qaw}.

While the number of GRBs detected in the VHE regime  in the afterglow increases, our understanding  of the physics underlying these  bursts remains superficial. Do GRBs with VHE emission share common properties? Can we use VHE observations  to infer properties of the CBM? 
In this paper, we intend to infer the characteristic features of GRBs exhibiting VHE emission and explore whether these bursts occur in  environments with similar properties, possibly different from the ones expected from typical Wolf-Rayet stars observed in our Galaxy~\citep[e.g.][]{2011AIPC.1358..165S, Crowther:2006dd}. 

This paper  is organized as follows. Section~\ref{sec:events} presents an  outline of the properties of the  bursts detected in the VHE regime. In Sec.~\ref{sec:standardAfterglow}, we review the afterglow model, the blaswave dynamics and the related synchrotron radiation. 
 We present constraints on the GRB energetics and the initial Lorentz factor in Sec.~\ref{sec:vhe}, while constraints on the non-observation of neutrinos from these bursts are presented in Sec.~\ref{sec:neutrinos}. A discussion on our findings is reported in Sec.~\ref{sec:discussion}, before concluding in Sec.~\ref{sec:conclusions}.
 The modeling of the photon energy distribution is summarized in Appendix~\ref{appedix0}, while 
 the  physics of hadronic interactions is outlined in Appendix~\ref{appedix1}. 
  Appendix~\ref{sec:comparison} provides additional insight on the properties of the CBM for our VHE GRB sample in comparison with GRBs without VHE emission.

\section{Sample of  gamma-ray bursts  observed at very high energies}\label{sec:events}
 The GRBs detected with VHE emission can  be broadly grouped in two classes 
 based on  the isotropic energy emitted in gamma-rays during the prompt phase: GRBs with intermediate to low isotropic energy [$\tilde{E}_\mathrm{\gamma, \rm{iso}} \lesssim 10^{50}$~erg, i.e. GRB 201015A~\citep{Suda:2021dox} and GRB 190829A~\citep{HESS:2021dbz}]  and energetic events with isotropic energy larger than typically observed  ($\tilde{E}_\mathrm{\gamma, \rm{iso}} \gtrsim 10^{53}$~erg). We limit our analysis to the latter group.

The class of bursts detected in the VHE regime and with large $\tilde{E}_{\gamma, \rm{iso}}$ is populated by: 
\begin{itemize}
\item[-] GRB 180720B detected with $\tilde{E}_{\gamma, \rm{iso}} \simeq 6 \times 10^{53}$~erg~\citep{2018GCN.23011....1F}. The H.E.S.S. Collaboration reported the observation of photons with energy between $0.11$~TeV and $0.44$~TeV at about $\sim 10$~hours after the trigger~\citep{Abdalla:2019dlr}.

\item[-] GRB 190114C, whose isotropic energy is estimated to be  $\tilde{E}_{\gamma, \rm{iso}} \simeq 2.5 \times 10^{53}$~\citep{2019GCN.23707....1H}. MAGIC observed $0.3$--$1$~TeV photons~\citep{MAGIC:2019irs, MAGIC:2019lau} from this burst, starting approximately one minute after its trigger. From Fig.~3 of \cite{MAGIC:2019lau}, we infer that a large number of photons of energy up to $1$~TeV is still observed at late times, around $520$~s, when the emission can be 
associated with the afterglow.

\item[-] GRB 221009A observed  with isotropic energy $\tilde{E}_{\gamma, \rm{iso}} \gtrsim 3 \times 10^{54}$~erg~\citep{2022GCN.32648....1D,2022GCN.32668....1F, 2022GCN.32762....1K}.   This is an interesting burst  with  photons with energy up to $18$~TeV  reported by LHAASO within $2000$~s post GBM trigger~\citep{2022GCN.32677....1H, 2022GCN.32748....1X}.  The distribution in energy and time of the VHE photons is not yet available,{while upper limits on the VHE flux at very late times have been published by H.E.S.S.~\citep{HESS:2023qhy}}.  On the contrary, the photon with energy $\simeq 0.4$~TeV detected  $0.4$~days after the trigger of the burst by \textit{Fermi}-LAT can be safely associated with the afterglow emission~\citep{2022GCN.32748....1X}.
\end{itemize}

Note that VHE emission has been observed from GRB 201216C as well, whose prompt isotropic energy is $\tilde{E}_{\gamma, \rm{iso}} \simeq 4.7 \times 10^{53}$~erg~\citep{2020GCN.29084....1F}. Since the published data is sparse to date~\citep{MAGIC:2021jcq},  we do not consider this GRB in our analysis.  The properties of the sample of GRBs that we consider throughout this paper are summarized  in Table~\ref{tab:sample}. 
\begin{table*}
    \caption{Properties of the sample of VHE GRBs  considered in this work. For each GRB we list the  redshift ($z$), the inferred emitted isotropic energy of the prompt phase  in the source frame ($\tilde{E}_\mathrm{\gamma, {iso}}$), the duration of the prompt emission ($T_{90}$), the time of detection of the VHE photon ($T_{\gamma, \rm{VHE}}$), the electron spectral index ($k_e$) and  the CBM type (wind or ISM). The following references are quoted in the table: [1]~\protect\cite{2018GCN.22981....1R}, [2]~\protect\cite{2018GCN.23011....1F}, [3]~\protect\cite{Abdalla:2019dlr}, [4]~\protect\cite{MAGIC:2019irs}, [5]~\protect\cite{MAGIC:2019lau}, [6]~\protect\cite{2019GCN.23688....1G}, [7]~\protect\cite{MAGIC:2019irs}, [8]~\protect\cite{Misra:2019vdg}, [9]~\protect\cite{2022GCN.32677....1H}, [10]~\protect\cite{2022GCN.32748....1X}, [11]~\protect\cite{2022GCN.32642....1L}, [12]~\protect\cite{2022GCN.32645....1B}.}
    \label{tab:sample}
    \centering
    \begin{tabular}{ccccccccc}
    Event & Redshift & $\tilde{E}_\mathrm{\gamma, iso}$ [erg] &   $T_{90}$ [s] & $T_{\rm{VHE}}$ [days] & $k_e$ &  CBM  & References  \\\hline 
    GRB 180720B & 0.653 & $6 \times 10^{53}$ & $49$ & $0.5$ & $2.4$ &  ISM &   [1, 2, 3]  \\
    GRB 190114C & 0.4245 & $2.5 \times 10^{53}$ & $25$ & $6 \times 10^{-3}$ &  $2.2$--$2.45$ &  ISM & [4, 5, 6, 7, 8]  \\
    GRB 221009A & 0.151 & $3 \times 10^{54}$ &  $300$ & $0.3$  &$2.5$ & Wind  &  [9, 10, 11, 12]\\
    \end{tabular}
\end{table*}

\section{Afterglow model}\label{sec:standardAfterglow}
 
In this section, we review the dynamics of the GRB blastwave as it propagates in the CBM. We also  introduce the synchrotron spectrum invoked to model the standard afterglow emission.

\subsection{Blastwave dynamics}
Throughout the prompt phase, the Lorentz factor of GRB outflows is $\Gamma \gg 100$~\citep{Gehrels:2009qy}. During the afterglow,  $\Gamma^{-1} \lesssim \theta_j$, being $\theta_j$ the jet half-opening angle. Therefore, it is safe to model the afterglow radiation through isotropic equivalent quantities~\citep{Kumar:2014upa}. We introduce the kinetic isotropic energy of the blastwave, $\tilde{E}_{k, \rm{iso}} = \tilde{E}_{\rm{iso}} - \tilde{E}_{\gamma, \rm{iso}}$, corresponding to the energy left in the outflow after the isotropic energy $\tilde{E}_{\gamma, \rm{iso}}$ has been released in gamma-rays during the prompt emission.

In the standard picture, the onset of the afterglow coincides with the beginning of the blastwave deceleration, occurring as the mass swept-up from the CBM becomes comparable to the initial mass of the outflow~\citep[e.g.,][and references therein]{zhang_2018}. The CBM is assumed to have particle density profiles scaling as $n \propto R^{-k}$, where $R$ is the distance from the central engine. Two asymptotic scenarios are usually considered in the literature~\citep{2011AIPC.1358..165S}: $k=0$, corresponding to a constant density interstellar medium (hereafter named ISM), and $k=2$, corresponding to a wind-like CBM (hereafter dubbed wind). 

As the blastwave expands, it interacts with the cold CBM. Two  shocks form: the forward shock, which propagates in the cold CBM, and the reverse shock, propagating in the relativistic jet, in mass coordinates. We focus on the self-similar phase, starting when the reverse shock has crossed the ejecta and the electromagnetic emission is mainly due to the forward shock. In this phase, the blastwave dynamics is well described by the Blandford-McKee  (BM) solution~\citep{1976PhFl...19.1130B}. 

The deceleration time of the blastwave depends on the particle density profile of the CBM. Assuming that the outflow is launched with initial Lorentz factor $\Gamma_0$ in the ISM  scenario~\citep{1976PhFl...19.1130B, zhang_2018}: 
\begin{equation}
    T_{\rm{dec, ISM}} = \left[ \frac{3 \tilde{E}_{k, \rm{iso}} (1+z)^3}{64 \pi n_0 m_p c^5 \Gamma_0^8} \right]^{1/3} \ ,
    \label{eq:tdecISM}
\end{equation}
where $n= n_0$ is the ISM density, $z$ is the redshift of the source, $c$ is the speed of light, and $m_p$ is the proton mass. 
As for the wind scenario, the number density of the CBM is parametrized as $n = A R^{-2}$. Here, $A= \dot{M}_w/( 4 \pi v_w m_p) = 3.02 \times 10^{35} A_\star$~cm$^{-1}$, where $A_\star = \dot{M}_{-5}/v_8$ is given for the typical mass loss rate $\dot{M}_{-5} = \dot{M}/( 10^{-5} M_\odot \; \rm{yr}^{-1})$ and wind velocity $v_8 = v_w/( 10^8 \; \rm{cm} \; \; \rm{s}^{-1})$ of Wolf-Rayet stars~\citep{Chevalier:1999jy, Razzaque:2013dsa}. According to this~\citep{Chevalier:1999mi}:
\begin{equation}
    T_{\rm{dec, wind}}  =  \frac{\tilde{E}_{k, \rm{iso}} (1+z)}{16 \pi A m_p c^3 \Gamma_0^4} \ .
    \label{eq:tdecWIND}
\end{equation}

After the deceleration starts, the Lorentz factor of the blastwave decreases with time~\citep{1976PhFl...19.1130B, Sari:1997qe, Chevalier:1999mi}:
\begin{eqnarray}
    \Gamma_{\rm{ISM}}= \Gamma_0  \left( \frac{T_{\rm{dec, ISM}}}{4 t}\right)^{3/8} \ , \label{eq:gammaISM} \\
    \Gamma_{\rm{wind}}= \Gamma_0 \left(  \frac{T_{\rm{dec, wind}}}{4 t}\right)^{1/4} \ \label{eq:gammaWIND}, 
\end{eqnarray}
for the ISM and wind scenarios, respectively.

Finally, the radius of the blastwave evolves as~\citep{Razzaque:2013dsa}:
\begin{equation}
    R= \frac{\zeta \Gamma^2 t c}{(1+z)} \ ,
    \label{eq:radius}
\end{equation}
where $\Gamma$  decreases with time according to  Eqs.~\ref{eq:gammaISM} or \ref{eq:gammaWIND}, and we recall that the time $t$ is measured in the observer frame. The parameter  $\zeta$ depends on the hydrodynamics of the blastwave. It is usually assumed to be constant, but its value is very uncertain~\citep[e.g.,][]{Sari:1997qe,Waxman:1997yv, Dai:1998iz, Derishev:2021ivd, Razzaque:2013dsa}; throughout this work, we adopt $\zeta=8$~\citep{Razzaque:2013dsa}.

We assume the uniform shell approximation of the BM solution. This is a fair assumption, since we are not interested in the hydrodynamics of the blastwave. Furthermore, the particle density of the BM shell quickly drops outside the region of width $\propto R/ \Gamma^2$ behind the forward shock. Hence, particle emission from outside this region is negligible. 

\subsection{Synchrotron spectrum}
As the fireball expands in the cold CBM, the forward shock at its interface converts the kinetic energy of the blastwave into internal energy, whose density is given by~\citep{1976PhFl...19.1130B}
\begin{equation}
    u^\prime= 4 m_p c^2 n \Gamma \left( \Gamma -1 \right) \ ,
    \label{eq:en_density}
\end{equation}
where $n= n_0$ for the ISM scenario and $n= A R^{-2}$ in the wind scenario.  Equation~\ref{eq:en_density} directly follows from the shock-jump conditions at the forward shock.

A fraction $\varepsilon_B$ of the internal energy density in Eq.~\ref{eq:en_density} is stored in the magnetic field, whose comoving strength is
\begin{equation}
    B^{\prime}= \sqrt{32 \pi m_p c^2 n \varepsilon_B \Gamma (\Gamma -1)} \ . 
    \label{eq:magneticField}
\end{equation}

The forward shock driven by the ejecta into the CBM is collisionless, meaning that it is mediated by collective plasma instabilities rather than collisions~\citep{Levinson:2019usn}. Hence, it can accelerate particles through the Fermi mechanism~\citep{Waxman:1995vg, Vietri:1995hs, Waxman:1998tn}. 
In particular, we assume that electrons are  accelerated to a power-law distribution $N(\gamma_e) \propto \gamma_e^{-k_e}$, where $k_e$ is the electron spectral index. The resulting non-thermal population of accelerated electrons is assumed to 
carry a fraction $\varepsilon_e$ of the energy density (Eq.~\ref{eq:en_density}). 

Three characteristic Lorentz factors  
define the distribution of shock-accelerated electrons: the minimum ($\gamma^\prime_{e, \rm{min}}$), the cooling ($\gamma^{\prime}_{e, \rm{cool}}$), and the maximum ($\gamma^\prime_{e, \max}$) ones. These are given by~\citep{Piran:2004ba, Chevalier:1999mi, Panaitescu:2000bk}:
\begin{eqnarray}
 \gamma^\prime_{e, \rm{min}} &=& \frac{\epsilon_e}{\xi_e} \frac{m_p}{m_e} {\frac{(k_e - 2)}{(k_e - 1)} (\Gamma - 1)}\ , \label{eq:gamma_min} \\ 
 \gamma^\prime_{e, \rm{cool}} &=& \frac{6 \pi m_e c}{\sigma_T B^{\prime 2}} \frac{(1+z)}{t \Gamma}\ , \label{eq:gamma_cool}\\
 \gamma^\prime_{e, \rm{max}} &=& \biggl( \frac{6 \pi e}{\sigma_T B^\prime \varphi} \biggr)^{1/2}\ ,\label{eq:gammaSat}
\end{eqnarray}
where $\sigma_T$ is the Thompson cross section, $\xi_e$ is the fraction of accelerated electrons, $e= \sqrt{\alpha \hbar c}$ is the electron charge, with $\alpha=1/137$ being the fine-structure constant, and $\hbar$ the reduced Planck constant. Finally, $\varphi$ is the number of gyroradii required to accelerate particles~\citep{2012JCAP...11..058G}. The maximum Lorentz factor $\gamma^\prime_{e, \max}$ is obtained by equating the electron cooling time $t^\prime_{e, \rm{cool}} = 6 \pi m_e c/( \sigma_T \gamma^\prime_e B^{\prime 2})$ and the acceleration time $t^\prime_{\rm{acc}} = 2 \pi \gamma^\prime_e m_e c^2 \varphi /(e c B^\prime)$.

The synchrotron break frequencies in Eqs.~\ref{eq:gamma_min}--\ref{eq:gammaSat} should take into account SSC losses of electrons, usually modeled through a correction factor depending on the Comptonization parameter $Y$~\citep[for more details, see \textit{e.g.}][]{Sari:2000zp}. For all  considered GRBs, observations show that the flux normalizations in the X-ray and VHE bands are comparable, hinting that synchrotron and SSC processes equally contribute to the cooling of electrons at the time of VHE emission. Since the $Y$ parameter decreases with time~\citep{Sari:2000zp}, and our analysis mainly considers epochs $t > T_{\rm{VHE}}$, we can safely neglect SSC corrections in Eqs.~\ref{eq:gamma_min}-\ref{eq:gammaSat}; see Sec.~\ref{sec:multiwave}. 

The  characteristic Lorentz factors of electrons introduce three energy breaks in the observed spectrum of synchrotron photons, namely $E_{\gamma, \min}$, $E_{\gamma, \rm{cool}}$ and $E_{\gamma, \rm{max}}$, defined as~\citep{Sari:1997qe}:
\begin{equation}
E_{\gamma} \equiv h \nu_{\gamma} = \frac{3}{2} \frac{B^\prime}{B_Q} m_e c^2 \gamma^{\prime\ 2}_e \frac{\Gamma}{(1+z)}\ ,
\label{eq:synch_energies}
\end{equation}
where $B_Q = 4.41 \times 10^{13}$~G.

The synchrotron self-absorption (SSA) Lorentz factor should be included for a complete treatment of synchrotron radiation. The corresponding break frequency is  expected in the radio band~\citep{zhang_2018}. However, detailed knowledge on the thermal electron distribution and on the structure of the emitting shell is needed to account for the SSA process~\citep{Warren:2018lyx}. We neglect this characteristic Lorentz factor and corresponding break frequency and discuss how this choice  affects our findings in Sec.~\ref{sec:compactness}. 

Electrons can be in two distinct radiative regimes: the ``fast cooling regime''  (if $\nu_{\gamma, \rm{min}} > \nu_{\gamma, \rm{cool}}$) or the ``slow cooling regime'' (for $\nu_{\gamma, \rm{min}} < \nu_{\gamma, \rm{cool}}$). In the former case, all the electrons efficiently cool down via synchrotron to the cooling Lorentz factor $\gamma_{e, \rm{cool}}$. In the latter case, synchrotron cooling is inefficient and it takes place  for electrons with $\gamma_{e} > \gamma_{e, \rm{cool}}$ only.

In the fast cooling regime, the synchrotron photon energy density [in units of GeV$^{-{1}}$~cm$^{-3}$] is~\citep{Sari:1997qe}: 
\begin{equation}
    n^{\prime \rm{sync}}_{\gamma}(E^\prime_\gamma) = A^\prime_{\gamma}
    \begin{system}
    \left( \frac{E^\prime_\gamma}{E^\prime_{\gamma, \rm{cool}}} \right)^{-\frac{2}{3}} \; \; \; \; \; \; \;\;\; \; \; \; \; \; \; \; \; E^\prime_\gamma< E^\prime_{\gamma, \rm{cool}}\\
\left( \frac{E^\prime_\gamma}{E^\prime_{\gamma, \rm{cool}}} \right)^{-\frac{3}{2}}  \; \;\;\; \; \;\;   \; \; \; \; \; \; \; \; \; \;   E^\prime_{\gamma, \rm{cool}} \leq E^\prime_\gamma \leq E^\prime_{\gamma, \rm{min}} \\
\left( \frac{E^\prime_{\gamma, \rm{min}} }{ E^\prime_{\gamma, \rm{cool}}} \right)^{-\frac{3}{2}} \left(\frac{E^\prime_\gamma}{E^\prime_{\gamma, \rm{min}}}\right)^{-\frac{k_e+2}{2}} E^\prime_{\gamma, \rm{min}} < E^\prime_\gamma \leq E^\prime_{\gamma, \rm{max}}
    \end{system}
    \ .
    \label{eq:lum_fast}
\end{equation}
On the other hand, in the slow cooling regime, the synchrotron photon energy density is:
 \begin{equation}
    n^{\prime \rm{sync}}_{\gamma}(E^\prime_\gamma)  =  A^{\prime}_\gamma
    \begin{system}
\left(\frac{E^\prime_\gamma}{E^\prime_{\gamma, \min}} \right)^{-\frac{2}{3}} \; \;\; \; \; \; \; \; \; \; \;\; \; \; \; \; \; \; \; \; \; \;\;  E^\prime_\gamma < E^\prime_{\gamma, \min}\\
\left(\frac{E^\prime_\gamma}{ E^\prime_{\gamma, \min}}\right)^{-\frac{(k_e+1)}{2}} \; \; \; \; \; \; \; \; \;  \; \;\; \; \;\;\;  E^\prime_{\gamma, \min} \leq E^\prime_\gamma \leq E^\prime_{\gamma, \rm{cool}} \\
\left(\frac{E^\prime_{\gamma, \rm{cool}}}{E^\prime_{\gamma, \rm{min}}} \right)^{-\frac{k_e+1}{2}} \left(\frac{E^\prime_\gamma}{E^\prime_{\gamma, \rm{cool}}} \right)^{-\frac{k_e+2}{2}}   E^\prime_{\gamma, \rm{cool}} <  E^\prime_\gamma \leq E^\prime_{\gamma, \rm{max}}
\end{system}
\ .
\label{eq:lum_slow}
\end{equation}
The normalization constant is given by~\citep{Sari:1997qe, Dermer:2000yd} 
\begin{equation}
A^{\prime}_\gamma= \frac{L^{\prime}_{\gamma, \rm{max}}}{4 \pi R^2 c \; \min(E^{\prime}_{\gamma, \min}, {E^{\prime}_{\gamma, \rm{cool}}})} \ ,
\end{equation}
where $L^{\prime}_{\gamma, \rm{max}}= N_e P^\prime_{\max}(\gamma^\prime_{e})/E^\prime_{\gamma}$ is the comoving specific luminosity [in units of s$^{-1}$]. 
The total number of radiating electrons in the blastwave is $N_e = 4 \pi n_0 \xi_e R^3/3$ in the ISM scenario, while it is given by $N_e = 4 \pi A \xi_e R $ in the wind scenario. Finally, the synchrotron power radiated by the electrons with Lorentz factor $\gamma^{\prime}_{e}= \min(\gamma^\prime_{e, \min}, \gamma^\prime_{e, \rm{cool}})$ is  $P^\prime_{\max}(\gamma^\prime_{e}) = c \sigma_T B^{\prime\ 2}\gamma^{\prime\ 2}_{e} /{(6 \pi)}$.

Given the photon energy density in Eqs.~\ref{eq:lum_fast} and \ref{eq:lum_slow}, the photon synchrotron spectrum observed at Earth  is [in units of GeV cm$^{-2}$ s$^{-1}$ Hz$^{-1}$]:
\begin{equation}
    \Phi^{\rm{sync}}_{\nu_\gamma}(E_\gamma; z)= \frac{(1+z)^2}{4 \pi d_L(z)^2} n^{\prime\ \rm{sync}}_{\gamma}\left( \frac{E_\gamma (1+z)}{\Gamma}\right) \frac{1}{\nu_\gamma} \frac{V^\prime_\gamma (1+z) }{t \Gamma}  \ \label{eq:photonFlux},
\end{equation}
where $V^\prime_\gamma = 4 \pi R^3/ 8 \Gamma$ is the comoving emitting volume of the blastwave and $d_L(z)$ is the luminosity distance of the source  at redshift $z$.
We assume a flat $\Lambda$CDM cosmology with 
$H_0 =  67. 4$~km~s$^{-1}$~Mpc$^{-1}$, $\Omega_M = 0.315$, and $\Omega_\Lambda = 0.685$~\citep{ParticleDataGroup:2020ssz}. 
The modeling of  the (V)HE spectrum complementing the synchrotron one is described in Appendix~\ref{appedix0}.

\section{Constraints on the energetics and initial Lorentz factor} \label{sec:vhe}
\begin{table*}
    \centering
    \caption{Multi-wavelength fluxes used in our analysis for the bursts listed in Table~\ref{tab:sample}. For each GRB, we list the considered observation time ($T_{\rm{obs}}$), as well as the correspondent radio flux ($F_{\nu, \rm{rad}}^{\rm{obs}}$), optical flux ($F_{\nu, \rm{opt}}^{\rm{obs}}$), and X-ray flux ($F_{\nu, \rm{X}}^{\rm{obs}}$); each at its  corresponding frequency or bands, as specified in parenthesis. The following references are quoted in the table: [1]~\protect\cite{2018GCN.23037....1S}, [2]~\protect\cite{Fraija:2019whb}, [3]~\protect\cite{Swift}, [4]~\protect\cite{Misra:2019vdg}, [5]~\protect\cite{Ren:2022icq}, [6]~\protect\cite{2022GCN.32655....1F}; see also references therein for the extrapolated fluxes.}
    \begin{tabular}{cccccc}
         Burst & T$_{\rm{obs}}$~[days] &$F_{\nu, \rm{rad}}^{\rm{obs}}$~[Jy]  & $F_{\nu, \rm{opt}}^{\rm{obs}}$~[Jy] &  $F_{\nu, \rm{X}}^{\rm{obs}}$~[Jy] & References \\ \hline
         GRB 180720B & $2$ & $10^{-3} \; (15.5$~GHz) &  $4 \times 10^{-5} \; (R$-band) & $1.24 \times 10^{-7} \; (10$~keV) & [1, 2, 3]\\
         GRB 190114C & $1.424$ & $1.930 \times 10^{-3} \; (5.5$~GHz) & $3.9 \times 10^{-5} \; (R-$band) & $5.98 \times 10^{-8} \; (10$~keV) & [3, 4] \\ 
         GRB 221009A & $2.3$ & $9 \times 10^{-3} \; (6$~GHz) & $2.016 \times 10^{-3} \; (R-$ band) & $2.19 \times 10^{-6} \; (10$~keV) & [3, 5, 6]\\ 
    \end{tabular}
    \label{tab:fluxes}
\end{table*}
In this section, we  present constraints on the blastwave energy and the surrounding CBM properties by exploiting the observed radio, optical and X-ray fluxes, and the opacity to $\gamma$--$\gamma$ pair production. By combining the observation of VHE photons with the duration of the prompt emission, we also infer upper and lower limits on the initial Lorentz factor $\Gamma_0$. We stress that we rely on the standard afterglow model outlined in Sec.~\ref{sec:standardAfterglow}. Hence, our constraints hold within this framework only. 

Among the GRBs listed in Table~\ref{tab:sample}, we select GRB 221009A and GRB 190114C to carry out our analysis. These GRBs are the closest ones and we consider them as  representative of our sample in terms of energetics, see Sec.~\ref{sec:events} and Table~\ref{tab:sample}. Furthermore, they are good examples of the  main models invoked to explain the VHE emission: SSC for GRB 221009A~\citep{Ren:2022icq} and proton synchrotron for GRB 190114C~\citep{Isravel:2022glo}. The parameters listed in Table~\ref{tab:sample} are fixed in our analysis, while we consider $\tilde{E}_{k, \rm{iso}}$, $n$, $\varepsilon_e$, and $\varepsilon_B$ as free parameters in the model.

\subsection{Multi-wavelength observations} \label{sec:multiwave}
As discussed in Sec.~\ref{sec:standardAfterglow}, the dynamics of the blastwave is independent of the initial Lorentz factor $\Gamma_0$,  and it is completely determined by the isotropic kinetic energy $\tilde{E}_{k, \rm{iso}}$ and the CBM density $n$. Hence, by requiring that Eq.~\ref{eq:photonFlux} matches the  fluxes observed across different wavebands, we can constrain the allowed $\tilde{E}_{k, \rm{iso}}$ and $n$. 

For GRB 221009A and GRB 190114C, the radio, optical and X-ray fluxes are extracted at the observation time $T_{\rm{obs}}$ where the data in the three wavebands are available. $T_{\rm{obs}}$ considered for each burst and the corresponding observed fluxes are listed in Table~\ref{tab:fluxes}. Multi-wavelength light-curves and tables of data are provided in~\cite{Misra:2019vdg} for GRB 190114C and in~\cite{Ren:2022icq} for GRB 221009A; see also references therein for observations with different instruments. The X-ray fluxes are obtained from the~\cite{Swift}. 

{We assume that the evolution of the emitting blast-wave is adiabatic, and that the micro-physical parameters of the emission are constant with time. Note that a different choice of $T_{\rm{obs}}$ would lead to the same order of magnitude estimation that we present here for $\tilde{E}_{k, \rm{iso}}$ and $n$. For convenience, we carry our analysis out at $T_{\rm{obs}}$ when radio, optical and X-ray data are simultaneously available for each GRB; see Table~\ref{tab:fluxes}.}

\begin{figure*}
    \centering
    \includegraphics[width=0.44\textwidth]{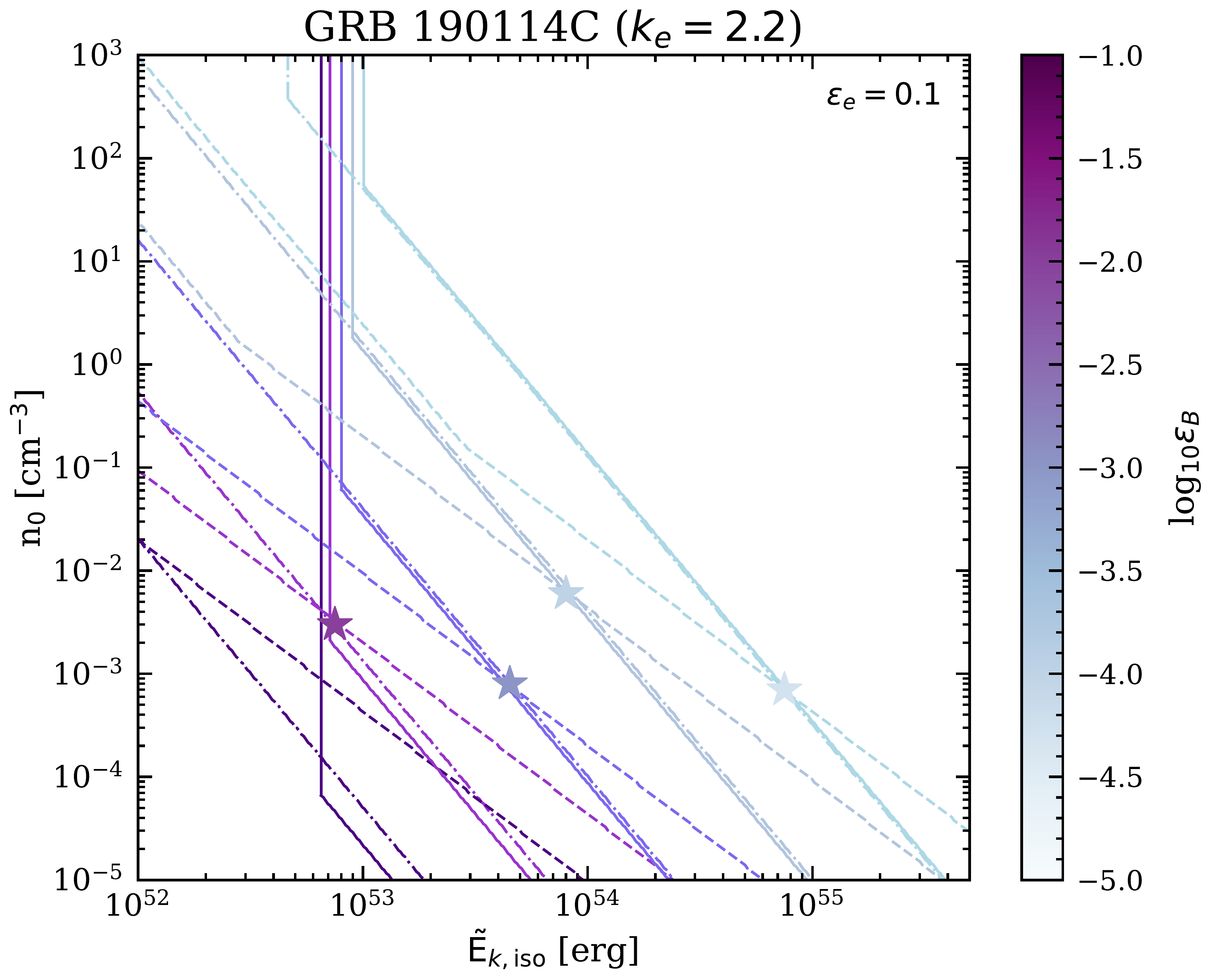}
    \includegraphics[width=0.44\textwidth]{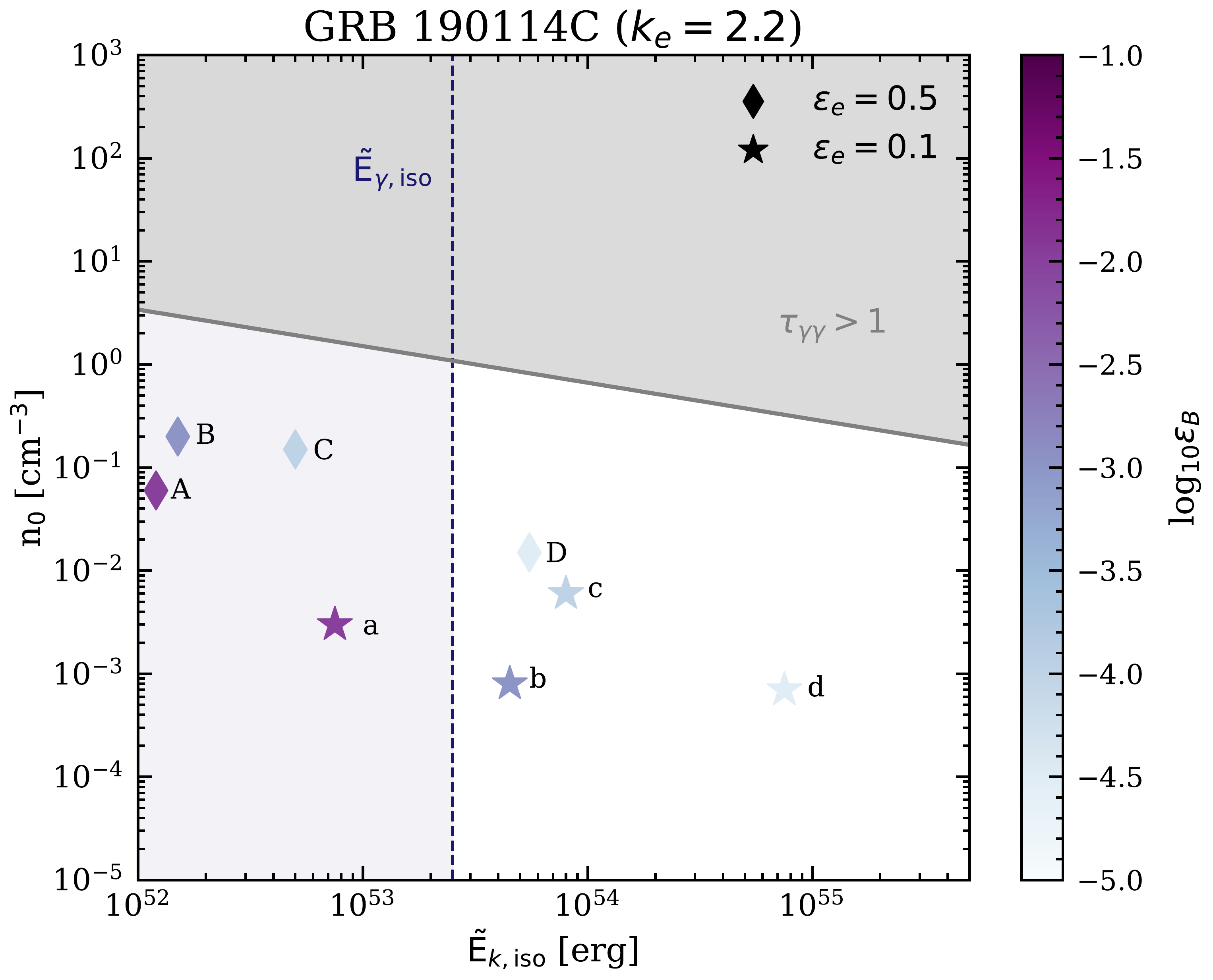}
    \includegraphics[width=0.44\textwidth]{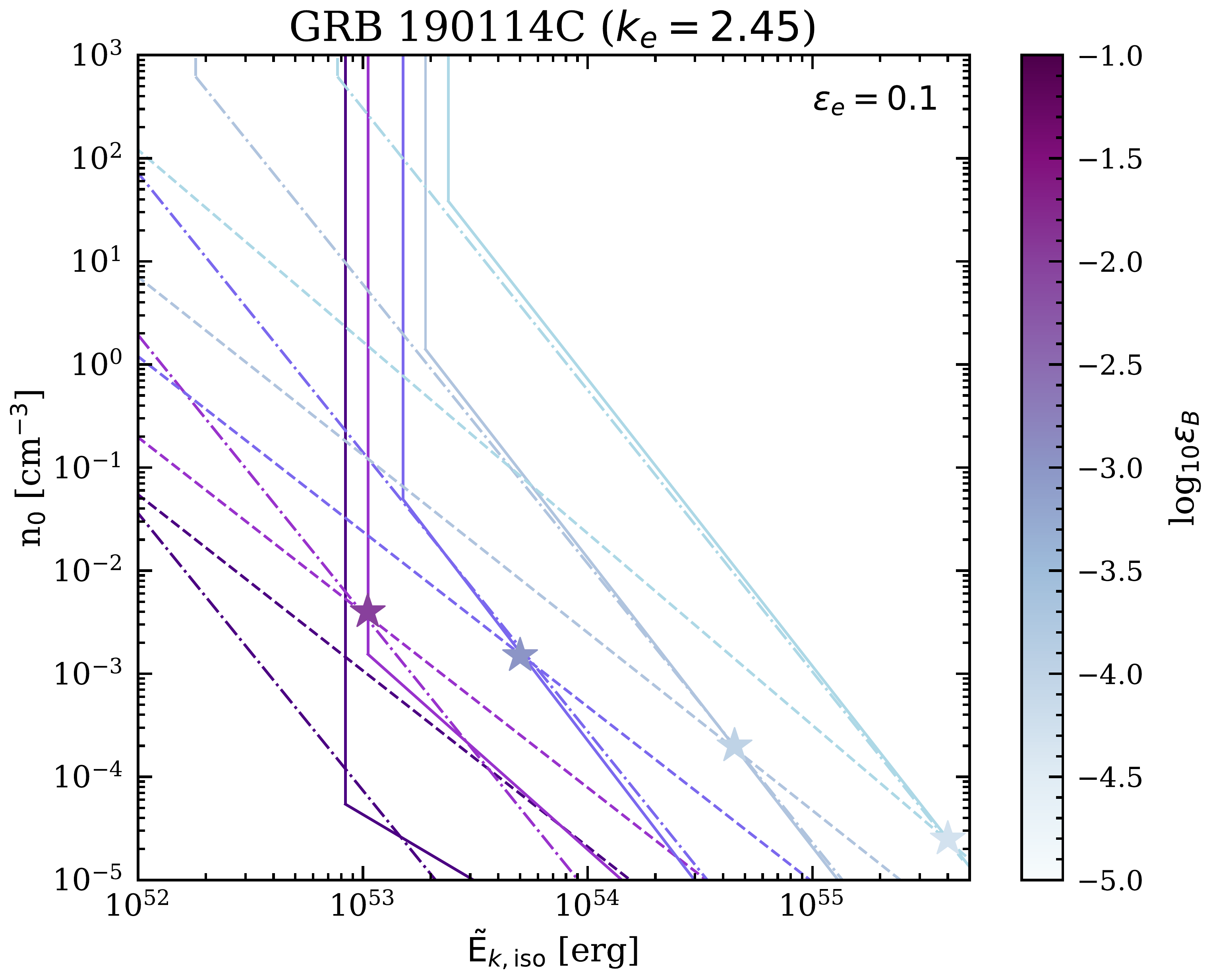}
    \includegraphics[width=0.44\textwidth]{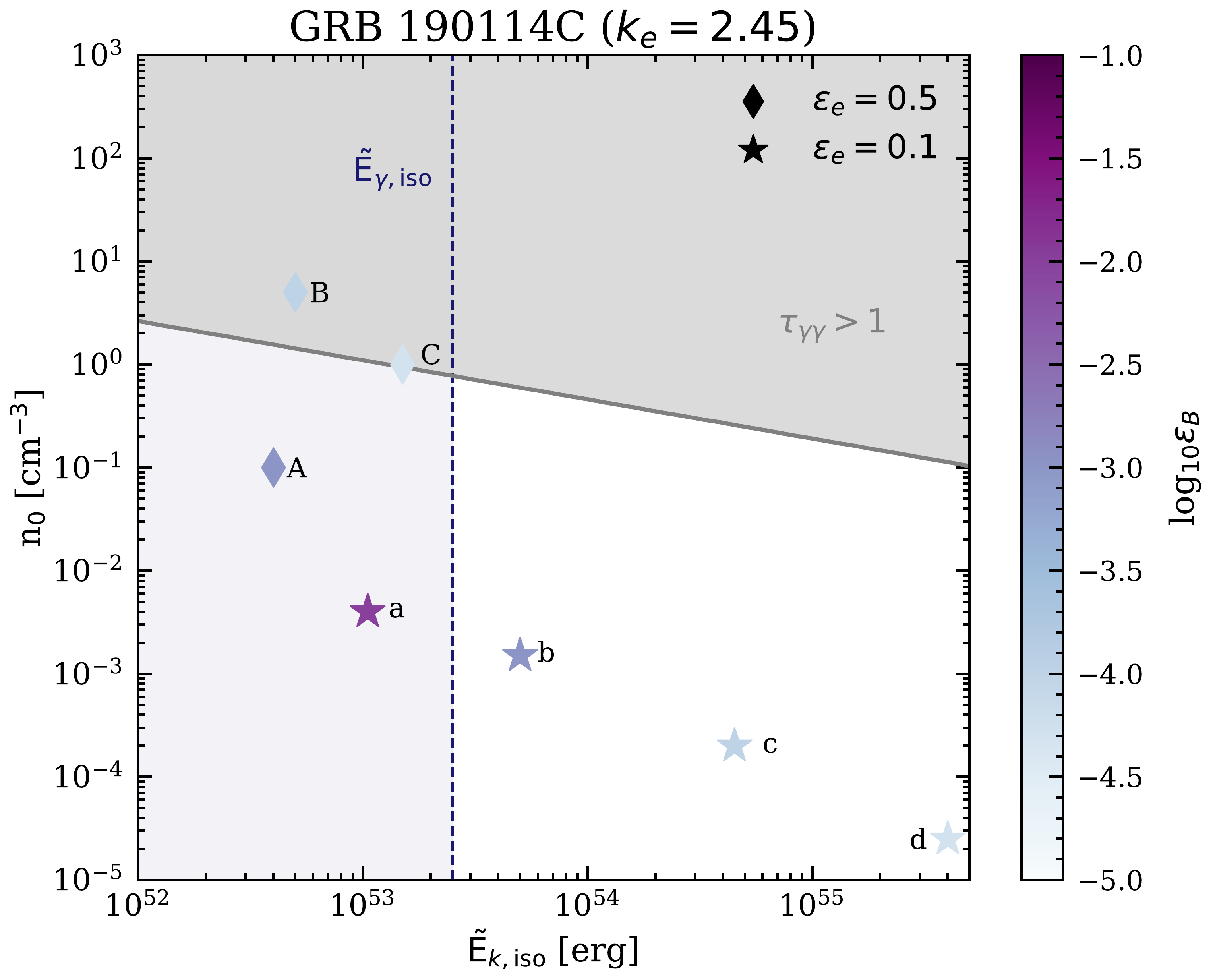}
    \includegraphics[width=0.44\textwidth]{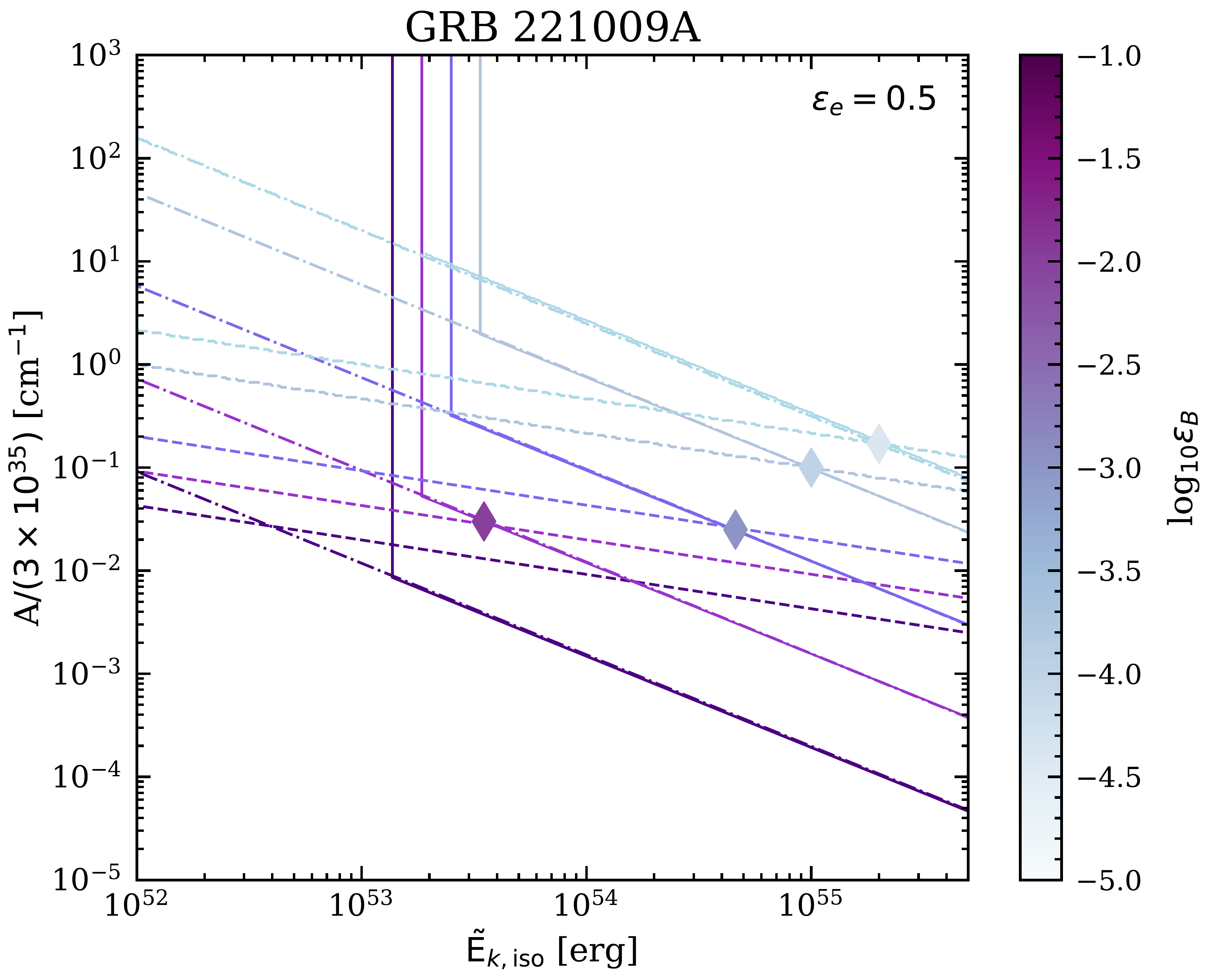}
    \includegraphics[width=0.44\textwidth]{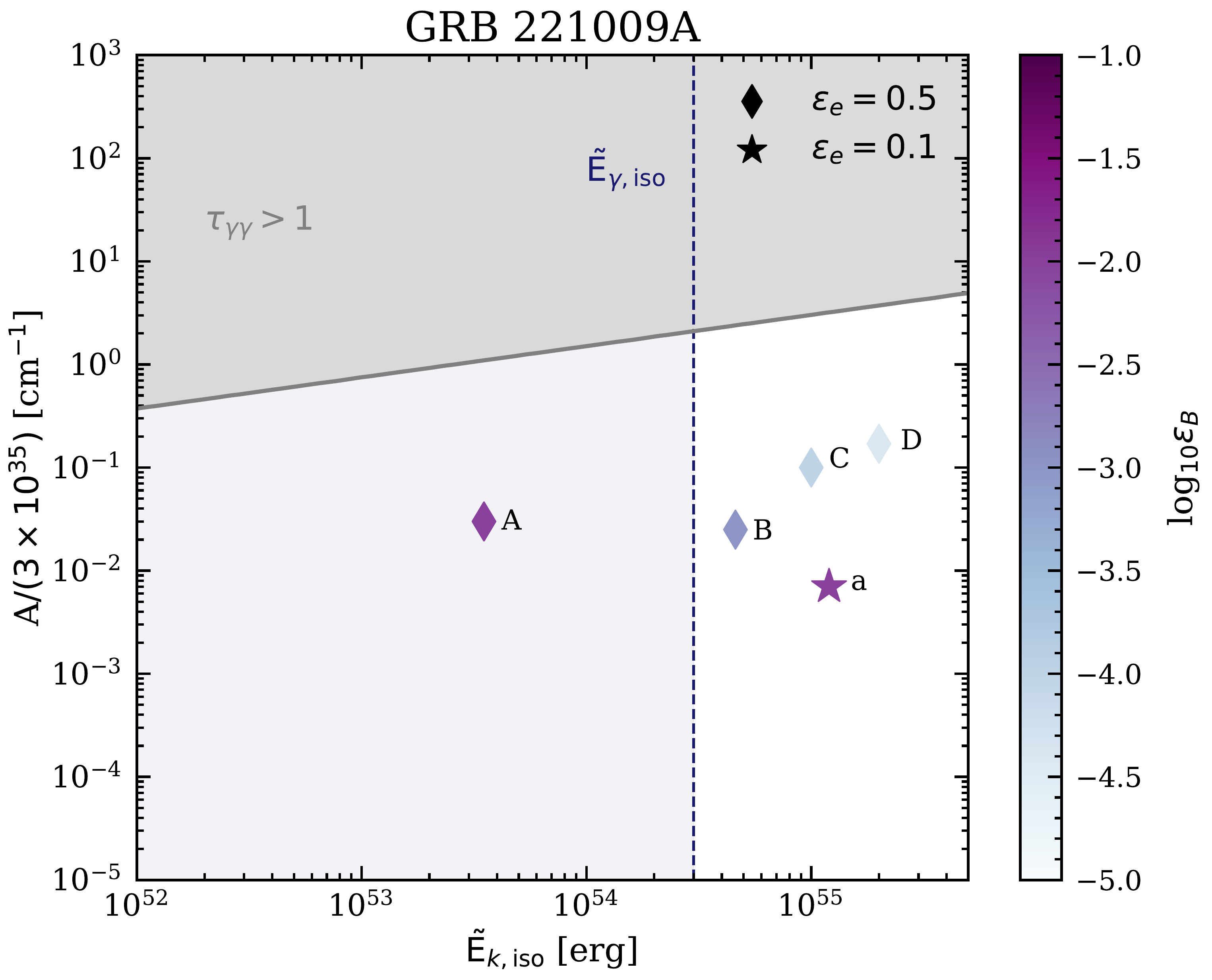}
    \caption{Isotropic kinetic energy $\tilde{E}_{k, \rm{iso}}$ and density $n_0 \;[A/(3 \times 10^{35})]$ compatible with the radio, optical and X-ray fluxes  for GRB 190114C (top and middle panels) and GRB 221009A (bottom panels). \textit{Left panels}:  $(\tilde{E}_{k, \rm{iso}}, n)$ pairs for which the computed synchrotron flux (Eq.~\ref{eq:photonFlux}) matches  the observed one in the radio (dashed lines), optical (dotted-dashed lines) and X-ray (continuous lines), as listed in Table~\ref{tab:fluxes}. For each burst, the line colors are defined by the value of $\varepsilon_B \in [10^{-5}, 10^{-1}]$ (see colorbar). For GRB 190114C, we fix $\varepsilon_e=0.1$ and $k_e=2.2$ in the top panels, while $k_e=2.45$ in the middle panels. For  GRB 221009A, $\varepsilon_e=0.5$ and $k_e=2.6$; see main text for details. The intersection among the lines for each value of $\varepsilon_B$ is marked by a star ($\varepsilon_e=0.1$) or a diamond ($\varepsilon_e=0.5$). Results are shown for the value of $\varepsilon_e$ which guarantees  solutions for $\varepsilon_B$ in the considered range. \textit{Right panels}: Same as the left panels, but  highlighting  the pairs of $(\tilde{E}_{k, \rm{iso}}, n)$ that simultaneously match radio, optical and X-ray data both for $\varepsilon_e=0.1$ (stars) and $\varepsilon_e=0.5$ (diamonds). The shadowed gray region is excluded from the transparency argument, i.e.~$\tau_{\gamma \gamma} >1$ at $T_{\rm{VHE}}$ (Table~\ref{tab:sample}). The dashed blue line marks the value of $\tilde{E}_{\gamma, \rm{iso}}$ for both bursts. Combining the transparency argument, the typical prompt emission efficiencies, and multi-wavelengths data, the preferred region of the parameter space for GRB 190114C [GRB 221009A] is the one with $ 2.5 \times 10^{53} \lesssim \tilde{E}_{k, \rm{iso}} \lesssim  10^{55}$~erg [$ 3 \times 10^{54} \lesssim \tilde{E}_{k, \rm{iso}} \lesssim  5 \times 10^{55}$~erg] and $ 6 \times 10^{-4} \lesssim n_0 \lesssim 2 \times 10^{-2}$~cm$^{-3}$ [$ 7 \times 10^{-3} \lesssim A/(3 \times 10^{35}) \lesssim  10^{-1}$~cm$^{-1}$]. Note that the upper limit set for the kinetic energy is implied by the requirement that $\varepsilon_B$ cannot be too small in collisionless shocks. The letters  mark the selected $(\tilde{E}_{k, \rm{iso}}, n)$ pairs for which the corresponding  initial Lorentz factor $\Gamma_0$ is reported in  Table~\ref{tab:gamma0}.}
    \label{fig:constraints}
\end{figure*}

The left panels of Fig.~\ref{fig:constraints} display the pairs of  $(\tilde{E}_{k, \rm{iso}}, n)$  for which  Eq.~\ref{eq:photonFlux} reproduces the fluxes observed  in the radio, optical and X-ray bands, respectively, for GRB 190114C (top and middle panels) and GRB 221009A (bottom panel).
For GRB 190114C we calculate the theoretical synchrotron flux for two values of the electron spectral index: $k_e= 2.2$ (top panels), which is obtained by inspecting the spectral energy distribution~\citep{Isravel:2022glo}, and $k_e=2.45$ (middle panels), which instead reproduces the slope of the lightcurve~\citep{Misra:2019vdg}. 
For GRB 221009A, we only consider $k_e=2.6$~\citep{Ren:2022icq}; 
see Table~\ref{tab:sample}. In all cases, we fix $\xi_e=1$ throughout our analysis. 
The line colors in the left panels of Fig.~\ref{fig:constraints} correspond to different values of $\varepsilon_B$, which we vary in the range $10^{-5}$--$10^{-1}$. 
For each $\varepsilon_B$ we select a value of $\varepsilon_e$ that allows for solutions, namely  $\varepsilon_e=0.1$ (stars) and $\varepsilon_e=0.5$ (diamonds) for GRB 190114C and GRB 221009A, respectively.
The intersection among the three lines in each of the left panels of Fig.~\ref{fig:constraints},  marked by a star (diamond), corresponds to the values of $\tilde{E}_{k, \rm{iso}}$ and $n$ which simultaneously reproduce the observed flux across the three wavebands for given pairs of ($\varepsilon_e$, $\varepsilon_B$). 

{The choice $\xi_e=1$ naturally excludes the proton synchrotron process for the modeling of the VHE emission, whereas it is consistent with the SSC scenario. The latter also requires $\varepsilon_e \gtrsim \varepsilon_B $, with typical parameters being $\varepsilon_e \simeq \mathcal{O}(10^{-1})$ and $\varepsilon_B \lesssim \mathcal{O}(10^{-2})$~\citep[e.g.,][]{Sari:2000zp}; the relation between $\varepsilon_e$ and $\varepsilon_B$ is inverted in the proton synchroton scenario, that is to say $\varepsilon_e \ll \varepsilon_B$~\citep[e.g.][]{Razzaque:2009rt, Isravel:2022glo}. Our assumptions are thus consistent with the SSC interpretation of the VHE emission. We discuss how this may affect our results in the following; see Sec.~\ref{sec:compactness} and Sec.~\ref{sec:discussion}.}

Note that we neglect any uncertainty on  the observed fluxes and the microphysical parameters for simplicity, and the lines in the left panels of Fig.~\ref{fig:constraints} are obtained by considering nominal values for the involved quantities.  
Furthermore, we rely on two approximations. First, we do not consider the exact hydrodynamics of the blastwave and adopt the uniform BM shell dynamics, as outlined in Sec.~\ref{sec:standardAfterglow}. Second, our results are sensitive to the constant $\zeta$ appearing in the definition of the blastwave radius, i.e.~Eq.~\ref{eq:radius}. 
However, we expect the error introduced by these two approximations to be below a factor of $2$.
Hence, the results in Fig.~\ref{fig:constraints}, although approximated,  provide  good insights into the features of our VHE GRB sample, if the standard afterglow model is adopted to explain multi-wavelength data. 

\subsection{Blastwave opacity to $\gamma$--$\gamma$ pair production}\label{sec:compactness}
The  synchrotron model, outlined in Sec.~\ref{sec:standardAfterglow} and adopted in Sec.~\ref{sec:multiwave}, cannot explain the VHE radiation observed during the afterglow, if the energy cutoff of relativistic electrons is taken into account~\citep{HESS:2021dbz}. Nevertheless, the energy cutoff cannot be neglected, and it is not clear under which conditions electrons can be accelerated up to PeV energies within the blastwave.  

To model the VHE emission, SSC has been invoked~\citep{Ghisellini:1998jy, Chiang:1998wh, Dermer:1999eh, Sari:2000zp, 2009ApJ...703..675N, Liu:2013foa, Asano:2020grw, Derishev:2021ivd, Fraija:2019rag} or mechanisms involving either proton-synchrotron radiation or the decay of secondaries produced in photo-pion and photo-pair processes~\citep[e.g.,][]{Bottcher:1998qi, Asano:2008tc, Razzaque:2009rt, Gagliardini:2022rrq, Isravel:2022glo}. 
Both these scenarios assume that the photons observed with $\sim$~TeV energy are produced in the same decelerating fireball as the synchrotron ones~\citep{1976PhFl...19.1130B, Sari:1997qe, Peer:2004doy}. Hence, in order to allow for VHE photons to escape the production region~\citep[e.g.,][]{Baring:1997am, Lithwick:2000kh}, the blastwave should  be transparent to $\gamma$--$\gamma$ pair production for $\mathcal{O}(\rm{TeV})$ photons for $t \gtrsim T_{\rm{VHE}}$,  $T_{\rm{VHE}}$ being the detection time of the VHE photon~\citep[e.g.,][]{Baring:1997am, Lithwick:2000kh}.

The blastwave opacity to $\gamma$--$\gamma$ annihilation is parameterized through the 
$\gamma$--$\gamma$ optical depth:
\begin{equation}
    \tau_{\gamma}(E^{\prime}_{\gamma, {\rm{VHE}}}) \simeq 0.1 \sigma_T E^\prime_\star n^{\prime \rm{sync}}_\gamma (E^\prime_\star) \frac{R}{\Gamma} \lesssim 1 \; ,
    \label{eq:tauGG}
\end{equation}
where $E^{\prime}_{\star}= (2 m_e c^2)^2/ E^{\prime}_{\gamma,{\rm{VHE}}}$, $E^\prime_{\gamma, {\rm{VHE}}}$ is the energy of the detected VHE photon, $R/ \Gamma$ is the compactness of the blastwave, and $n^{\prime \rm{sync}}_{\gamma}$ is the energy density of synchrotron photons (see Eqs.~\ref{eq:lum_fast} and \ref{eq:lum_slow})~\footnote{In principle, the whole  photon energy distribution, including the VHE component, should be used. Nevertheless, $E^\prime_\star$ falls between the optical and X-ray bands for the VHE photons we are interested in. Hence, in order to simplify the calculation, it is safe to consider the synchrotron component only.}. Note that Eq.~\ref{eq:tauGG}  evaluates the blastwave opacity at the peak of the $\gamma$--$\gamma$ annihilation cross section~\citep[see e.g.][]{2012MNRAS.421..525H}. 

As mentioned in Sec.~\ref{sec:multiwave}, the dynamics of the blastwave only depends on its isotropic kinetic energy $\tilde{E}_{k, \rm{iso}}$ and on the CBM density. Therefore, Eq.~\ref{eq:tauGG} further constrains the  $(\tilde{E}_{k, \rm{iso}}, n)$ pairs allowing  VHE photons to escape from the blastwave, independently on the model adopted for explaining the VHE emission.  

The right panels of Fig.~\ref{fig:constraints} show the region of the parameter space that does not fulfill Eq.~\ref{eq:tauGG} at $T_{\rm{VHE}}$ and for the observed $E_{\gamma, \rm{VHE}}$, both different for each burst (see Table~\ref{tab:sample}). In addition, a summary of the constraints obtained by combining the {radio, optical and X-ray} observations discussed in Sec.~\ref{sec:multiwave} is also displayed. {We stress that we do not aim to fit the multi-wavelength data and we do not include VHE fluxes in Fig.~\ref{fig:constraints}. Rather, we only require that the VHE photon escapes the blastwave at $T_{\rm{VHE}}$, according to Eq.~\ref{eq:tauGG}; this argument is different  than the one adopted in Sec.~\ref{sec:multiwave}. As already discussed in Sec.~\ref{sec:multiwave}, the choice $T_{\rm{obs}} = T_{\rm{VHE}}$ would not change the results in Fig.~\ref{fig:constraints}. Since the radio data are not available  at $T_{\rm{VHE}}$ for all the bursts,  we list the observed time when radio, optical and X-ray data are simultaneously available in Table~\ref{tab:fluxes}.}

In Fig.~\ref{fig:constraints}, we show results for $\varepsilon_e=0.1$ and $0.5$, while we verified that smaller values of $\varepsilon_e$ do not allow to reproduce simultaneously the radio, optical and X-ray fluxes for any value of $\varepsilon_B$. This might depend on the fact that we neglect the SSA frequency, which would introduce an additional break in the photon distribution and shift the radio flux to larger values~\citep{Warren:2018lyx}. However, the considered bursts are expected to be in the weak-absorption regime at $T_{\rm{obs}}$ considered in our analysis, cf.~Table~\ref{tab:fluxes}~\citep{Fraija:2019wel, Fraija:2019whb}, while to date no information is available for GRB 221009A. Hence, neglecting the SSA process in the synchrotron spectrum may be a valid approximation.

{Our results depend  on  $\xi_e=1$. Smaller values of this parameter could allow $\varepsilon_e < 0.1$ and would lead to larger values of $n_0$ or $A/ ( 3 \times 10^{35} )$, typically inferred when the proton synchrotron model is adopted to explain the VHE emission, see e.g.~\cite{Eichler:2005ug, Isravel:2022glo}. Therefore, the results in Fig.~\ref{fig:constraints} are consistent within the SSC scenario, but no longer hold in the proton synchrotron one, as previously discussed in Sec.~\ref{sec:multiwave}.
Given the large number of degeneracies in the afterglow model, we limit our discussion to the case with $\xi_e=1$ and leave a detailed investigation of the dependence of our findings on this assumption to future work.}

The transparency argument is particularly powerful for GRB 190114C when the spectral index $k_e=2.45$ is adopted. In this case, some pairs $(\tilde{E}_{k, \rm{iso}}, n_0)$ which reproduce the flux across different wavebands are excluded by the requirement that the VHE photons do not undergo $\gamma$--$\gamma$ pair-production  at $t \lesssim T_{\rm{VHE}}$. 

The allowed parameter space can be further constrained by considering 
the radiative efficiency of the prompt phase. Despite the latter being a topic of debate and potentially varying depending on the event, we here adopt a typical efficiency of $\simeq 10 \%$~\citep[e.g.][]{Beniamini:2016hzc}.
Since $\tilde{E}_{\gamma, \rm{iso}} \sim 2.5 \times 10^{53}$~erg [$\tilde{E}_{\gamma, \rm{iso}} \gtrsim 3 \times 10^{54}$~erg] for GRB 190114C [GRB 221009A], we  expect  the region of the parameter space with $ 2.5 \times 10^{53} \lesssim \tilde{E}_{k, \rm{iso}} \lesssim  10^{55}$~erg [$ 3 \times 10^{54} \lesssim \tilde{E}_{k, \rm{iso}} \lesssim  5 \times 10^{55}$~erg] and $ 6 \times 10^{-4} \lesssim n_0 \lesssim 2 \times 10^{-2}$~cm$^{-3}$ [$ 7 \times 10^{-3} \lesssim A/(3 \times 10^{35}) \lesssim  10^{-1}$~cm$^{-1}$] to be preferred,  as indicated by the dashed blue line in the right panels of Fig.~\ref{fig:constraints}. 

The inclusion of SSA in our treatment could shift the densities to larger values. However, as already mentioned, GRB 190114C may be in the weak-absorption regime at the considered time~\citep{Fraija:2019wel, Fraija:2019whb}. Our CBM densities for GRB 190114C are much smaller than the ones inferred in~\cite{Isravel:2022glo}, which finds $n_0 \simeq \mathcal{O}(10$--$100)$~cm$^{-3}$. This is due to our assumption $\xi_e=1$, whereas $\xi_e \simeq \mathcal{O}(10^{-2})$ is required in~\cite{Isravel:2022glo} in the context of the proton synchrotron model for the VHE emission.

As for GRB 221009A,  our results are consistent with the ones of~\cite{Ren:2022icq}, which obtains $A_\star=1.2 \times 10^{-2}$ for $\tilde{E}_{k, \rm{iso}}=6.8 \times 10^{54}$~erg, $\varepsilon_e=0.2$ and $\varepsilon_B=2 \times 10^{-3}$. On the contrary, for GRB 190114C, we obtain $n_0 \lesssim 2 \times 10^{-2}$~cm$^{-3}$, which is a factor $\mathcal{O}(10)$ smaller than $n_0 = 0.3$~cm$^{-3}$ obtained in~\cite{Wang:2019zbs}. This discrepancy may be due to the fact that~\cite{Wang:2019zbs} does not take into account data in the radio band. As it can be seen in the left-middle panel of Fig.~\ref{fig:constraints}, when only the optical and X-ray fluxes are used, we recover $n_0 \simeq \mathcal{O}(10^{-1})$~cm$^{-3}$, if we assume $\tilde{E}_{k, \rm{iso}} = 6 \times 10^{53}$~erg, $\varepsilon_e=0.1$ and $\varepsilon_B=10^{-4}$, i.e.~for parameters compatible with the ones adopted in~\cite{Wang:2019zbs}. Thus, more solutions are possible if the radio data are not included in the analysis since the optical and X-ray data are degenerate for a large part of the $(\tilde{E}_{k, \rm{iso}}, n_0)$ space. 

For  GRB 180720B,  the compactness argument is not constraining. In fact,  a signal in the energy range $E_{\gamma, \rm{VHE}} = 0.11$--$0.44$~TeV has been reported for  this GRB at the time considered in Table~\ref{tab:sample}. At such late times, we expect the blastwave to be already transparent to $\gamma$--$\gamma$ pair production. Hence, we do not show plots for this burst. Nevertheless, 
exploiting the multi-wavelength data, our approach  enables us to break the degeneracies involved in the standard afterglow model and to obtain $6 \times 10^{53} \lesssim \tilde{E}_{k, \rm{iso}} \lesssim 10^{55}$~erg and $ 4 \times 10^{-5} \lesssim n_0 \lesssim 10^{-1}$~cm$^{-3}$. For this burst,  our parameters are similar to those inferred in~\cite{Wang:2019zbs}, namely $\tilde{E}_{k, \rm{iso}} = 10^{54}$~erg and $n_0=0.1$~cm$^{-3}$.

Additional inputs  on $\tilde{E}_{\gamma, \rm{iso}}$ may further restrict the allowed parameter space shown in Fig.~\ref{fig:constraints}, when typical prompt efficiencies are taken into account~\citep{Beniamini:2016hzc}. Our results hold if the multi-wavelength radiation observed from  this class of bursts is modelled within the standard afterglow framework outlined in Sec.~\ref{sec:standardAfterglow}. More complex jet geometries~\citep{Sato:2022kup}, time-varying microphysical parameters~\citep{Filgas:2011hj,Misra:2019vdg}, the assumption of two-zone models~\citep{Khangulyan:2023akp} or other more complex models~\citep[e.g.,][]{Laskar:2023yap} would affect our conclusions. {Intriguingly, a low-density wind environment is inferred for GRB 221009A in~\cite{Laskar:2023yap}, even though they suggest that the standard assumptions of the afterglow theory may be violated by this burst.}

\subsection{Initial Lorentz factor}\label{sec:gamma0}

As discussed in Sec.~\ref{sec:standardAfterglow}, the afterglow dynamics is  independent of the initial value of the blastwave Lorentz factor ($\Gamma_0$), if the shell is in the self-similar regime~\citep{1976PhFl...19.1130B, Sari:1997qe}. 
Therefore,  the afterglow onset (i.e., the deceleration time $T_{\rm{dec}}$) can be used to infer $\Gamma_0$. Assuming that the VHE  photon detected at $T_{\rm{VHE}}$ is associated with the afterglow, the blastwave should start to decelerate at $T_{\rm{dec}} \lesssim T_{\rm{VHE}}$. From Eqs.~\ref{eq:tdecISM} and \ref{eq:tdecWIND}, this  translates in a lower limit (LL) for $\Gamma_{0}$: 
\begin{eqnarray}
    \Gamma_{0, \rm{ISM}}^{\rm{LL}} &=& \left[ \frac{3 \tilde{E}_{k, \rm{iso}}(1+z)^3}{64 \pi n_0 m_p c^5 T_{ \rm{VHE}}^3} \right]^{1/8} \ , \\
    \Gamma_{0, \rm{wind}}^{\rm{LL}} &=& \left[ \frac{\tilde{E}_{k, \rm{iso}}(1+z)}{16 \pi A m_p c^3 T_{\rm{VHE}}} \right]^{1/4} \ ,
    \label{eq:lowerlimit}
\end{eqnarray}
for the ISM and wind scenarios, respectively. 

Even though there is no significant correlation between the onset of the afterglow $T_{\rm{dec}}$ and the duration of the prompt emission $T_{\rm{90}}$~\citep{Ghirlanda:2017opl}, the assumption of a thin shell---for which the reverse shock is at most mildly 
relativistic---implies $T_{\rm{dec}} \gtrsim T_{90}$. Within this approximation, most of the energy of the ejecta has been transferred to the blastwave at the onset of  deceleration~\citep{Hascoet:2013bma}. This condition provides us with  upper limits (UL) on $\Gamma_0$:
\begin{eqnarray}
    \Gamma_{0, \rm{ISM}}^{\rm{UL}} &=& \left[ \frac{3 \tilde{E}_{k, \rm{iso}}(1+z)^3}{64 \pi n_0 m_p c^5 T_{90}^3} \right]^{1/8} \ , \\
    \Gamma_{0, \rm{wind}}^{\rm{UL}} &=& \left[ \frac{\tilde{E}_{k, \rm{iso}}(1+z)}{16 \pi A m_p c^3 T_{90}} \right]^{1/4} \ ,
    \label{eq:upperlimit}
\end{eqnarray}
for the ISM and wind scenarios, respectively. 

For  fixed isotropic kinetic energy and CBM density, $\Gamma_{0, \rm{ISM (wind)}}^{\rm{LL}}$ can be obtained by rescaling $\Gamma_{0,\rm{ISM} (wind) }^{\rm{UL}}$ by $\left(T_{90}/ T_{\rm{VHE}} \right)^{3/8}$, if the burst propagates in a constant density medium, or by  $\left(T_{90}/ T_{\rm{VHE}} \right)^{1/4}$ in the wind scenario.  For each point marked in the right panels of Fig.~\ref{fig:constraints} through a letter, the  range of allowed values of $\Gamma_{0, {\rm{ISM (wind)}}} $ is listed in Table~\ref{tab:gamma0}.
\begin{table}
    \centering
    \caption{Upper and lower limits on $\Gamma_{0, \rm{ISM (wind)}}$ obtained for the points of the parameter space selected through the criteria illustrated  in the right panels of Fig.~\ref{fig:constraints} and marked by a letter therein. 
    }
    \begin{tabular}{cccc}
      Burst  & Symbol & $\Gamma_{0, \rm{ISM (wind)}}^{\rm{LL}}$ & $\Gamma_{0, \rm{ISM (wind)}}^{\rm{UL}}$ \\ \hline
       GRB 190114C ($k_e=2.2$) & a & 312 & 961 \\ 
         & b & 180 & 555 \\
         & c & 216 & 665 \\
    & d & 146 & 450 \\ 
    & A & 153 & 472 \\
     & B & 85 & 262 \\
     & C & 71 & 218 \\
    & D & 80 & 246 \\ \hline
    GRB 190114C ($k_e=2.45$) & a & 575 & 1797 \\
     & b & 337 & 1054 \\
     & c & 199 & 622 \\
     & d & 145 & 454 \\
    & A & 76 & 237 \\
     & B & 55 & 170  \\
     & C & 50 & 156 \\ \hline
    GRB 221009A & a & 173 & 313 \\
     & A & 50 & 160 \\
      & B & 47 & 154 \\
     & C & 55 & 180 \\
    & D & 27 & 90  
        \end{tabular}
    \label{tab:gamma0}
\end{table}

Our limits complement the estimates obtained from  the prompt emission for GRB 221009A~\citep{Murase:2022vqf, Liu:2022mqe, Ai:2022kvd}. Furthermore, they are in agreement with~\cite{Li:2022gat}, which obtains $\Gamma_0 = 719 \pm 59$ for GRB 190114C. Note that the lower limits $\Gamma_{0, \rm{wind}}^{\rm{LL}}$ for GRB 221009A are quite small and hence not constraining, due to the large $T_{\rm{VHE}}$ (see Table~\ref{tab:sample}). The results in Table~\ref{tab:gamma0} and Fig.~\ref{fig:constraints} hint that a very energetic blastwave propagating in a low density medium implies large $\Gamma_0$. This could be justified by considering that weaker winds extract less angular momentum from the GRB progenitors. In this scenario, the core collapse may be driven by faster rotation, which favors the formation of highly collimated jets, compatible with the large $\Gamma_0$ and isotropic energies in low-density CBMs~\citep{Hascoet:2013bma}. {Similar conclusions on the high collimation of GRB 221009A have been reached also in~\cite{Laskar:2023yap}.}

\section{Constraints from the non-observation of high-energy neutrinos}\label{sec:neutrinos}

Provided that protons are co-accelerated at the forward shock, the GRB afterglow is expected to emit neutrinos with  PeV--EeV energy~\citep{Waxman:1997ga, Dermer:2000yd, Li:2002dw, Razzaque:2013dsa, Murase:2007yt, Guarini:2021gwh}. Neutrinos are predominantly  produced through photo-hadronic ($p \gamma$) interactions of the protons accelerated at the external shock and photons produced as the blastwave decelerates, as summarized in Appendix~\ref{appedix1}. 

The IceCube Neutrino Observatory detects neutrinos in the TeV--PeV range~\citep{IceCube-Gen2:2021rkf, IceCube:2020abv}. Nevertheless, so far no neutrino detection has been reported in connection to electromagnetic observations of  GRBs~\citep{IceCube:2017amx}, with upper limits set on the prompt~\citep{IceCube:2021xar} and the afterglow emission~\citep{Lucarelli:2022ush, IceCube:2022rlk}. Yet, upcoming neutrino facilities, such as  IceCube-Gen2 and its radio extension~\citep{IceCube-Gen2:2021rkf}, the Radio Neutrino Observatory~\citep{RNO-G:2020rmc}, the  Giant Radio Array for Neutrino Detection (GRAND200k)~\citep{GRAND:2018iaj}, as well as the spacecraft Probe of Extreme Multi-Messenger Astrophysics (POEMMA)~\citep{Venters:2019xwi} are expected to improve the detection prospects of afterglow neutrinos. 

The non-observation of neutrinos from GRB 221009A~\citep{2022GCN.32665....1I} allows to constrain the GRB properties as well as the mechanism powering the prompt emission~\citep{Murase:2022vqf, Liu:2022mqe, Ai:2022kvd,Rudolph:2022dky}. We intend to  investigate whether  complementary constraints can be obtained through  the current non-detection of neutrinos from the afterglow of VHE GRBs.
To this purpose,  we  model the neutrino signal expected from the afterglow of  GRB 190114C, since  multi-wavelength interpretations invoking both SSC and proton synchrotron have been proposed~\citep{Wang:2019zbs, Isravel:2022glo}. {We focus on the SSC model, since the results outlined in Sec.~\ref{sec:compactness} are consistent with this interpretation, and  briefly discuss the proton synchrotron case.} We expect the correspondent neutrino signal to be representative for all other GRBs in our sample (Table~\ref{tab:sample}). However, more detections in the VHE band would allow to make more accurate predictions. 

The time-integrated neutrino signal from $p \gamma$ interactions is calculated  following  Sec.~4 of~\cite{Guarini:2021gwh}, using as input the total photon distribution (defined in Eq.~\ref{eq:totalPh}) and the proton distribution  (Eq.~\ref{eq:proton_distribution}). The parameters adopted for {computing the neutrino signal within} the SSC model are summarized in Table~\ref{tab:paramNeut} {, corresponding to~\cite{Wang:2019zbs}} \footnote{{Note that we rely on the findings of~\cite{Wang:2019zbs} only in this section, since their work performs a multi-wavelength fit including the VHE component. Our discussion in Sec.~\ref{sec:multiwave} is independent on~\cite{Wang:2019zbs}. The microphysical parameters obtained in~\cite{Wang:2019zbs} are consistent with ours, while the density $n_0$ is a factor $\mathcal{O}(10)$ larger than the one obtained in Sec.~\ref{sec:multiwave} for GRB 190114C. As a consequence, the neutrino signal presented  in this section is an upper limit with respect to the  one we would obtain using the results of  Sec.~\ref{sec:multiwave}.}}.
\begin{table}
    \caption{Assumed model parameters for GRB 190114C resulting from the multi-wavelength modeling of the photon distribution outlined in~\protect\cite{Wang:2019zbs}.}
    \label{tab:paramNeut}
   \center
    \begin{tabular}{cc}
    Parameter & SSC fit\\\hline 
    $\tilde{E}_{k, \rm{iso}}$~[erg] & $6 \times 10^{53}$ \\
    $n_0$~[cm$^{-3}$] & $0.3$  \\
    $\Gamma_0$ & $300$ \\
    $\xi_e$ & 1  \\
    $\xi_p$ & 1  \\
    $\varepsilon_e$ & $0.07$  \\
    $\varepsilon_B$ & $4 \times 10^{-5}$  \\
    $\varepsilon_p$ & $0.8$  \\ 
    $\varphi$ & 10 \\
    $k_e$ & $2.5$  \\
    $k_p$ & $2.2$ \\
    $\delta$ & $-26^\circ$ \\
    \end{tabular}
\end{table}
For protons we fix $\xi_p=1$---in order to obtain an optimistic estimation of the resulting neutrino flux--- and $\varepsilon_p = 1- \varepsilon_e -\varepsilon_B$ and $k_p=2.2$~\citep{Sironi:2013ri}. As a consequence, the  neutrino flux computed in the SSC scenario represents an upper limit to the actual flux for the considered $\tilde{E}_{k, \rm{iso}}$, since no constraints can be derived on the fraction of energy going into accelerated protons nor on the fraction of accelerated protons.

\begin{figure*}
\centering
\includegraphics[width=0.475\textwidth]{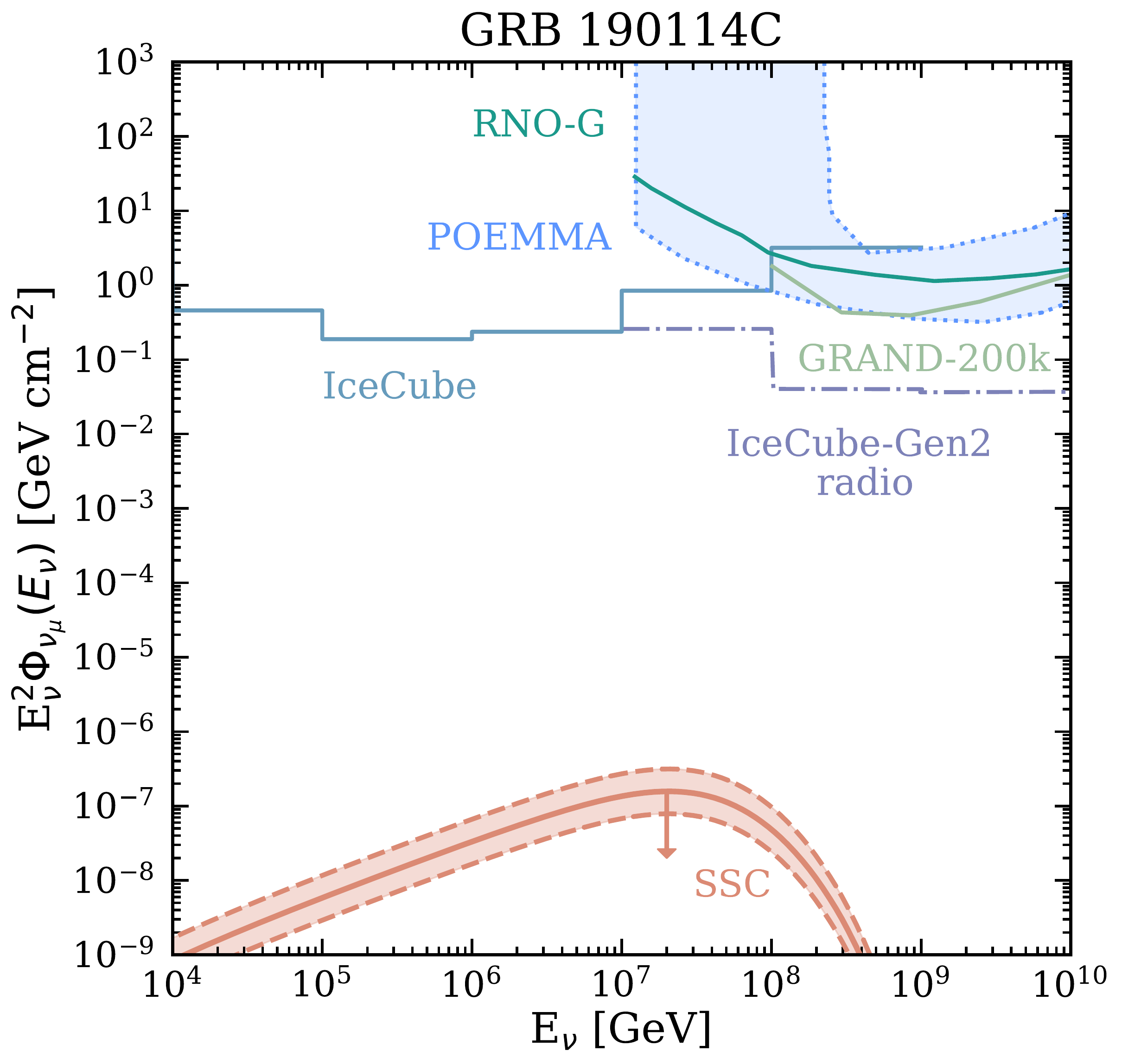}
\includegraphics[width=0.43\textwidth]{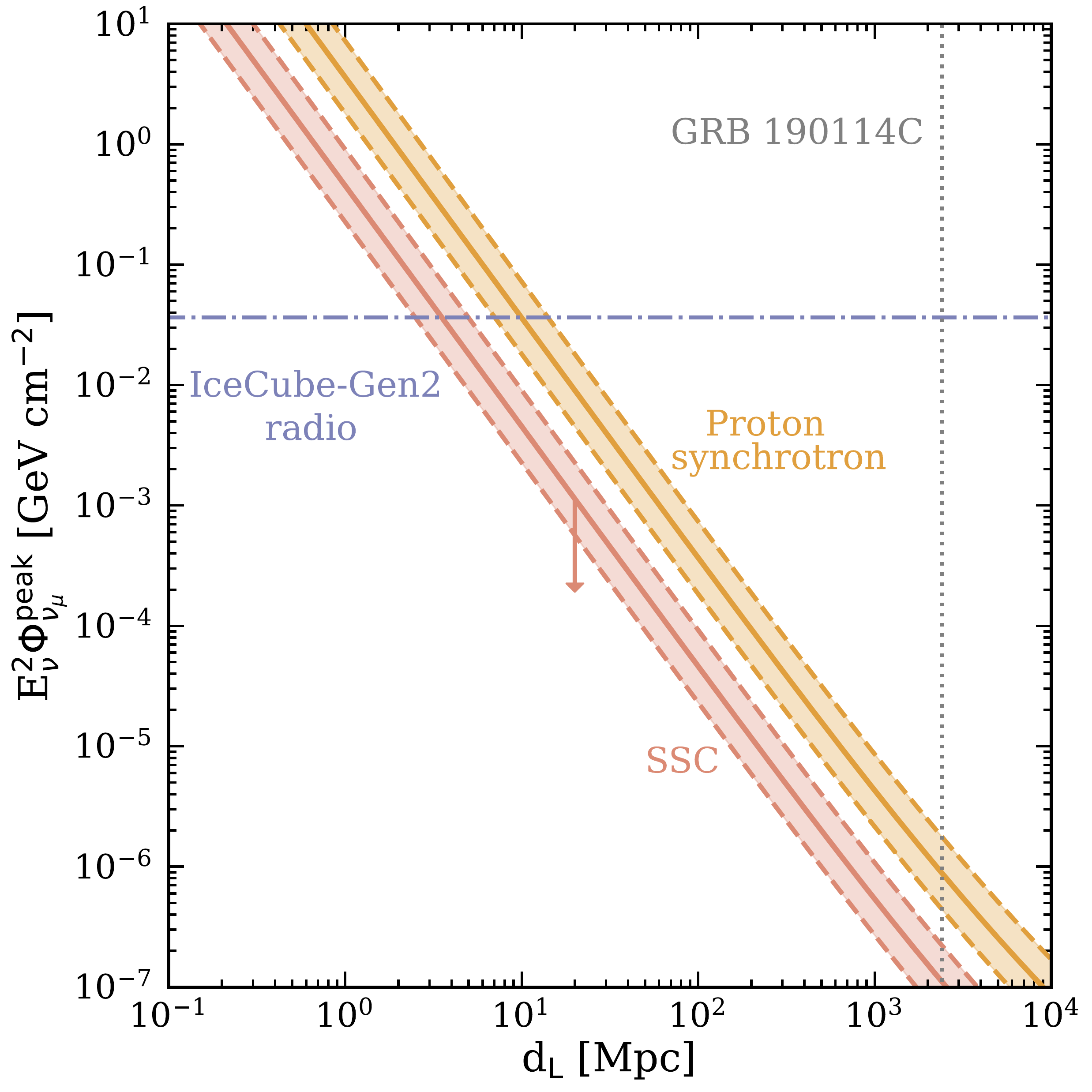}
\caption{ Time-integrated muon neutrino flux expected from the afterglow of GRB 190114C.  \textit{Left panel}: Time-integrated muon neutrino flux for the  SSC model (pink shadowed area). This  flux represents an upper limit for the real one (for the here considered $\tilde{E}_{k, \rm{iso}}$), as denoted  by the pink arrow. The error in the flux prediction is assumed to be a factor $\Delta \Phi_{\nu_\mu}/ \Phi_{\nu_\mu} = \pm 2$, due to the uncertainties in the analytical prescription of the photon flux; see main text. For comparison, the sensitivity of IceCube for a source located at the declination $\delta = -23^\circ$~\citep{IceCube-Gen2:2021rkf, IceCube:2020xks} is shown as well as the most optimistic ones of IceCube-Gen2 radio for a source at $\delta=0^\circ$~\citep{IceCube-Gen2:2021rkf}, RNO-G for $\delta=77^\circ$~\citep{RNO-G:2020rmc}, GRAND200k for a source at $\left| \delta \right|=45^\circ$~\citep{GRAND:2018iaj}, and the full-range time-integrated sensitivity of POEMMA~\citep{Venters:2019xwi}. The neutrino signal lies well below the sensitivity curves of current and upcoming VHE neutrino telescopes. Even though this result depends on the assumed parameters, we expect these conclusions to hold for different sets of parameters within current uncertainties. \textit{Right panel}: Peak of the time-integrated muon neutrino flux (plotted in the left panel) as a function of the luminosity distance for a bursts with the same properties of GRB 190114C, except for its distance. The arrow for the SSC model is the same as the left panel. {For comparison, we show the same result for the proton synchrotron model (yellow shadowed area), by relying on the parameters inferred in~\citet{Isravel:2022glo}.} We also plot the sensitivity of IceCube-Gen2 radio for the optmistic case of a source located at  $\delta = 0^\circ$~\citep{IceCube-Gen2:2021rkf}. Note that since the peak of the time-integrated neutrino flux occurs for $E_{\nu} \simeq 10^7$--$10^8$~GeV for all the redshifts, we approximate the sensitivity of IceCube-Gen2 radio to be constant. In particular, we take the minimum value of the sensitivity to get the most optimistic prediction. For comparison, the dotted grey line indicates GRB 190114C. The peak of the time-integrated neutrino flux becomes comparable to the sensitivity of IceCube-Gen2 radio for $d_L \lesssim 5$~Mpc ($d_L \lesssim 15$~Mpc) for the SSC model parameters (proton synchrotron).
{As expected, at each distance the neutrino signal in the case
of the proton synchrotron model is larger than the SSC one, since the former
naturally requires larger values of $n_0$ and $\tilde{E}_{k, \mathrm{iso}}$.}}
\label{fig:neutrino}
\end{figure*}

The left panel of Fig.~\ref{fig:neutrino} shows the  time-integrated muon neutrino flux, $\Phi_{\nu_\mu}$, from the afterglow GRB 190114C for  the SSC model. For comparison, we also show the sensitivity of IceCube to a  source located at the declination $\delta \simeq -23^\circ$~\citep{IceCube-Gen2:2021rkf, IceCube:2020xks}. 
In order to investigate future detection prospects,   we  plot the 
most optimistic sensitivity of IceCube-Gen2 radio for a  source at $\delta = 0^\circ$~\citep{IceCube-Gen2:2021rkf}, the one of RNO-G for a source at $\delta= 77^\circ$~\citep{RNO-G:2020rmc}, as well as the sensitivity of GRAND200k for a source at $\left| \delta \right|= 45^\circ$~\citep{GRAND:2018iaj} and the full-range time-integrated sensitivity of POEMMA~\citep{Venters:2019xwi}.
We consider an error band $\Delta \Phi_{\nu_\mu}/\Phi_{\nu_\mu} = \pm 2 $, according to the uncertainties intrinsic to the  analytical model, as discussed in Sec.~\ref{sec:compactness}. The main uncertainties come from the choice of parameters listed in Table~\ref{tab:paramNeut}.
Nevertheless, the neutrino signal  lies well below the sensitivity curves also for optimistic values of $\varepsilon_e$ and $\varepsilon_B$---see also~\cite{Guarini:2021gwh}. These findings imply that the non-detection of neutrinos from the afterglow of GRBs with VHE emission is expected and does not allow to further constrain the properties of the bursts. Conversely, detection of VHE neutrinos in coincidence with VHE GRB afterglows would be challenging to explain in the context of the standard afterglow model.

The right panel of Fig.~\ref{fig:neutrino} shows the peak of the time-integrated neutrino flux (plotted in the left panel) as a function of the luminosity distance, assuming a burst with properties identical to the ones of  GRB 190114C. {For comparison, we also show the peak of the time-integrated neutrino flux when the proton synchrotron model is assumed. We rely on the parameters inferred in~\cite{Isravel:2022glo}. We warn the reader that they are not comparable with the ones obtained in Sec.~\ref{sec:multiwave} due to our assumption $\xi_e=1$ and the requirement $\xi_e \ll 1$ for a proton synchrotron model. Hence, the main goal of the right panel of Fig.~\ref{fig:neutrino} is to assess whether the neutrino detection perspectives from VHE bursts depend on the selected model for the VHE emission. Since the proton synchrotron model naturally requires larger values of $n_0$ and $\tilde{E}_{k, \rm{iso}}$, the resulting neutrino flux is larger than in the SSC scenario.}

Comparing the  peak of the neutrino flux to  the sensitivity of IceCube-Gen2 radio~\citep{IceCube-Gen2:2021rkf}, which is expected to be the most competitive facility (see left panel), we obtain that  the peak of the  neutrino flux becomes comparable to the sensitivity of IceCube-Gen2 radio for $d_L \lesssim 5$~Mpc ($d_L \lesssim 15$~Mpc) for the SSC model parameters (proton synchrotron). Such distances are too small, considering the distribution of  long GRBs as a function of the redshift~\citep{2012ApJ...752...62J}. 
Therefore, we conclude that the detection of neutrinos from  GRB afterglows displaying VHE emission is  not a promising tool to infer GRB properties within a multi-messenger framework. Our conclusions are consistent with the ones of~\cite{Isravel:2022glo} for GRB 190114C, which finds that the photo-hadronic interaction rate accounts for inefficient energy extraction.

\section{Discussion}\label{sec:discussion}
\begin{table*}
    \centering
    \caption{Summary of the model parameter constraints derived in this work from the afterglow of GRB 180720B, GRB 190114C and GRB 221009A (see also Fig.~\ref{fig:constraints}). The   range preferred for  the kinetic isotropic energy ($\tilde{E}_{k, \rm{iso}}$), the CBM density ($n_0  \; [A/(3 \times 10^{35})]$), and the initial Lorentz factor ($\Gamma_{0, \rm{ISM(wind)}}$) is reported. }
    \begin{tabular}{cccc}
      GRB   & $\tilde{E}_{k, \rm{iso}}$ [erg] & $n_0$ or $A/(3 \times 10^{35})$ & $\Gamma_{0, \rm{ISM} (wind)}$ \\ \hline
      GRB 180720B &  $ 6 \times 10^{53}$--$10^{55}$ & $4 \times 10^{-5}$--$10^{-1}$~cm$^{-3}$ & $80-1276$ \\ 
       GRB 190114C &  $2.5 \times 10^{53}$--$10^{55}$ & $3 \times 10^{-5}$--$2 \times 10^{-2}$ cm$^{-3}$ & $50$--$1797$ \\
       GRB 221009A  & $ 3 \times 10^{54}$--$5 \times 10^{55}$ & $7 \times 10^{-3}$--$ 10^{-1}$ cm$^{-1}$ & $ \lesssim 313$ 
        \end{tabular}
    \label{tab:summary}
\end{table*}
Our constraints on the VHE GRB properties  are summarized in Table~\ref{tab:summary} for our benchmark bursts, GRB 180720B, GRB 190114C and GRB 221009A (see also  Fig.~\ref{fig:constraints}).  While our sample is small, such findings raise questions on the nature of the progenitors and the sites hosting VHE bursts, {if microphysical parameters compatible with the SSC scenario are assumed}. 

The initial Lorentz factor of our  VHE bursts falls within the  average  expected for GRBs, see Sec.~\ref{sec:gamma0} and e.g.~Secs.~5 and 6 of \cite{Ghirlanda:2017opl}. With the caveat that we have observed  VHE emission for a few bursts only, our results seem to suggest that these VHE GRBs  exhibit isotropic kinetic energy towards the higher tail of the distribution expected for GRBs, see e.g.~Fig.~19 of \cite{Poolakkil:2021jpc}. This result might be biased by the sensitivity of existing telescopes, as well as the viewing angle. In  the future, CTA may detect fainter bursts in the VHE regime, providing better insight on the population features and the fraction of GRBs with VHE emission.

As discussed in Sec.~\ref{sec:vhe}, VHE GRBs  might preferentially occur in a low-density CBM--independently on the microphysics of the shock, the compactness argument requires that $n_0 \lesssim 1$~cm$^{-3}$ and $A/(3 \times 10^{35}) \lesssim 1$~cm$^{-1}$. {If microphysical parameters typical of SSC radiation are adopted~\citep[e.g.,][]{Fraija:2022rtj}, even more stringent constraints are obtained.}
Intriguingly, we reach similar conclusions following the method outlined in~\cite{Gompertz:2018anr} that relies on the simplifying assumption that all the bursts can be modelled with the same set of microphysical parameters and have a prompt emission efficiency $\simeq 50 \%$. Following~\cite{Gompertz:2018anr}, we find that the VHE GRBs cluster in the low-density region of the parameter space [$n_0 (A /(3 \times 10^{35})) \lesssim 10^{-1}$~cm$^{-3}$ (cm$^{-1}$)]. On the contrary, the bursts not displaying VHE emission analyzed in~\cite{Gompertz:2018anr} are uniformly distributed in the $(\tilde{E}_{k, \rm{iso}}, n)$ space; we refer the interested reader to Appendix~\ref{sec:comparison} for additional details.

A CBM with low density is usually favored by the synchrotron closure relations~\citep[e.g.][]{Gao:2013mia}, that  are found not to be fullfilled for all  bursts; hence our findings  might  be affected by the simplifications intrinsic to  these relations.  Yet, these results are in agreement with the expectation that low-density environments  favor a transparent blastwave in the afterglow. Furthermore, larger densities may reprocess the VHE photons and emit electromagnetic radiation in other wavelengths. 
In addition, low density CBMs have been  associated to  long GRBs~\citep[e.g.][]{2002ApJ...571..779P, Gompertz:2018anr}.
For example, a wind with $A_\star \simeq 4 \times 10^{-3}$ has been inferred for GRB 130427A~\citep{Panaitescu:2013pga}, as a result of a multi-wavelength fit of the GRB lightcurve.  \cite{Panaitescu:2013pga} suggests that the weak wind could be a consequence of the  GRB progenitor being  hosted in a superbubble~\citep{Mirabal:2002wk, Scalo:2001rq}. Similarly, \cite{Hascoet:2013bma}  finds that the winds of some GRB progenitors are weaker than the ones observed for  Wolf-Rayet stars in our Galaxy ($A_\star \simeq 1$). This might be linked to the low metallicity of the progenitors~\citep{Vink:2001cg} and their host galaxies~\citep[e.g.][]{Perley:2013fh}, which is anyway still under debate~\citep[e.g.][]{Perley:2015hfa}. Low CBM densities may also be  caused by  reduced mass-loss rate at the time of the stellar collapse~\citep{Hascoet:2013bma}.  {Recently, \citet{Dereli-Begue:2022clf} studied bursts with a plateau phase in their afterglow. In order to explain this feature, a small wind density consistent with our findings and a small outflow Lorentz factor are required, the latter implying a lack of VHE (and even HE) emission for those bursts, which~\cite{Dereli-Begue:2022clf} argues is the case.}

Our results hold within the assumption that the multi-wavelength radiation is generated by the decelerating blastwave, whose dynamics is outlined in Sec.~\ref{sec:standardAfterglow}. Such conclusions may substantially change if more complex jet geometries~\citep{Sato:2022kup}, time-varying microphysical parameters~\citep{Filgas:2011hj} or two-zone models~\citep{Khangulyan:2023akp} should be invoked, as in the case of GRB 190114C~\citep{Misra:2019vdg}. Furthermore, low-density CBMs are obtained for $\xi_e=1$, whereas smaller fractions of accelerated electrons {naturally} lead to larger densities~\citep[e.g.][]{Isravel:2022glo}. 
{Thus, if a dense CBM should be inferred, e.g. via the SSA frequency in the radio band, it may hint towards a proton synchrotron model. In this sense, determining the CBM density can provide constraints on the mechanism powering the VHE emission.}
{Our results are based on the SSC scenarios, rather than proton synchrotron ones. Since the value of $\xi_e$ is largely uncertain,  an analysis of the dependence of our conclusions on this parameter  is left to future work. Additional input on these parameters may also come from numerical simulations of particle acceleration at the external shock.}

Future observations of GRBs in the VHE regime with CTA~\citep{Knodlseder:2020onx} will be crucial to pinpoint the mechanism powering the VHE emission during the afterglow. However, CTA might have better detection prospects for large CBM surrounding these bursts, as suggested in~\cite{Mondal:2022uvu}. The latter assumes a SSC origin of the VHE emission, although neglecting the cutoff introduced by $\gamma$--$\gamma$ pair production. The SSC efficiency largely depends on the Compton parameter, which is maximized for large blastwave energies and CBM densities. As a consequence, \cite{Mondal:2022uvu} obtains CBM densities larger than the ones we infer, since the transparency argument alone is sufficient to limit $n_0 \lesssim 1$~cm$^{-3}$ [$A/(3 \times 10^{35}) \lesssim 1$~cm$^{-1}$]. We stress that the relation used for the opacity argument (Eq.~\ref{eq:tauGG}) is approximate; therefore, detailed modeling of the energy cutoff and fit to the spectral energy distributions are required to draw robust conclusions from a larger burst sample. Yet, we do not expect the constraint $n < 1$~cm$^{-1}$ to change drastically.

\section{Conclusions}\label{sec:conclusions}
While the number of GRBs detected in the VHE regime during the afterglow will increase in the near future with the advent of CTA, our understanding of the mechanism powering the VHE emission is very preliminary. The standard synchrotron model, which well explains the afterglow data from the radio to the X-ray bands, cannot account for the emission of $\mathcal{O}$(TeV) photons detected at late times. 

In this paper, we focus on  GRB 180720B, GRB 190114C and GRB 221009A, with the goal to infer the properties of the blastwave and the burst environment.
By requiring that the plasma in the blastwave shell is transparent to 
$\gamma$--$\gamma$ pair production at the time of the observation of the VHE photons, we obtain that the CBM density should be $n_0 \lesssim 1$~cm$^{3}$ [$A/(3 \times 10^{35})$~cm$^{-1}$]. A tentative interpretation of the radio, optical and X-ray data hints towards even lower CBM densities, with $n_0 \lesssim \mathcal{O}(10^{-1})$~cm$^{-3}$ [$A/(3 \times 10^{35}) \lesssim \mathcal{O}(10^{-1})$~cm$^{-1}$], {if the microphysical parameters of the shock are taken to be consistent with SSC mechanism}. Furthermore, we obtain constraints on the initial Lorentz factor of the blastwave by requiring that the deceleration of the fireball starts before the observation of  VHE photons  and after the GRB prompt emission, finding  $10^{2} \lesssim \Gamma_{0} \lesssim 10^{3}$.  While the initial Lorenz factors are within average in the context of long GRBs, 
we find that (assuming a typical prompt-phase efficiency of $10 \%$) the kinetic blastwave energy is large, 
$\tilde{E}_{k, \rm{iso}} \gtrsim  \times \mathcal{O}(10^{54})$~erg (see also Table~\ref{tab:summary}). Albeit such large energies could be due to an observational bias towards detection efficiency. Whether these conclusions are generally valid for VHE  GRBs 
will be confirmed by future CTA observations.

Finally, we investigate the neutrino signal expected from the afterglow of VHE GRBs, focusing on  GRB 190114C as representative burst. The non-observation of high-energy neutrinos from VHE GRBs is consistent with our theoretical predictions. 
The detection prospects for high-energy neutrinos from VHE GRBs with upcoming neutrino telescopes are equally poor, except for bursts closer than $15$~Mpc. This suggests
that neutrinos from the GRB afterglow may not be promising messengers to unveil the properties of the VHE emitting bursts. 

Our findings hint at arising trends characterizing the properties of VHE GRBs, if the afterglow of these bursts can be modelled within the standard scenario. Additional data on bursts exhibiting VHE emission will shed light on the engine powering such transients and provide valuable insight  on the characteristics of their host environments.

\section*{Acknowledgements}
We are very grateful to Jochen Greiner for insightful discussions. This project has received funding from the  Villum Foundation (Project No.~37358), the Carlsberg Foundation (CF18-0183), the MERAC Foundation, the Deutsche Forschungsgemeinschaft through Sonderforschungsbereich
SFB~1258 ``Neutrinos and Dark Matter in Astro- and Particle Physics'' (NDM), and the European Research Council via the ERC Consolidator Grant No.~773062 (acronym O.M.J.). 

\section*{Data Availability}
Data can be shared upon reasonable request to the authors. 

\bibliographystyle{mnras}
\bibliography{main}

\appendix

\section{Photon energy distribution}\label{appedix0}
The total distribution of target photons is 
\begin{equation}
    n^{\prime \rm{tot}}_{\gamma}(E^\prime_\gamma)= n^{\prime \rm{sync}}_{\gamma} (E^\prime_{\gamma})+ n^{\prime \rm{VHE}}_{\gamma} (E^\prime_\gamma) \ ,
    \label{eq:totalPh}
\end{equation} 
where $n^{\prime \rm{sync}}_\gamma$ is the synchrotron component defined in Eqs.~\ref{eq:lum_fast} (including SSC corrections, see~\cite{Sari:2000zp}) and \ref{eq:lum_slow} and $n^{\prime \rm{VHE}}_{\gamma}$ is the VHE part of the photon energy distribution. 

We model the VHE component of the photon spectrum both with SSC radiatio.
The SSC component is obtained by following the prescription in~\cite{2013MNRAS.435.2520G}. We include the Klein-Nishina regime by introducing a cut-off in the photon spectrum at the Klein-Nishina energy~\citep{Wang:2019zbs}. {The latter, can be expressed as~\cite{Wang:2019zbs}:}.
{
\begin{equation}
E_{\rm{KN}}=
    \begin{system}
        0.3 \; \rm{TeV} \; 6 \frac{k_e-2}{k_e-1} \varepsilon_{e, -1} E_{54}^{1/4} n^{-1/4}_{-1} t_{2}^{-3/4} \; \; \; \gamma_{\rm{min}}> \gamma_{\rm{cool}} \; \\
        0.1 \; \rm{TeV} \frac{1}{1+Y} \varepsilon_{B, -2}^{-1} E_{54}^{-1/4} n_{-1}^{-3/4} t_{10 \; \rm{hr}}^{-1/4} \; \; \; \; \gamma_{\rm{min}}< \gamma_{\rm{cool}} \; ,
    \end{system}
\end{equation}
where $Y$ is the Compton parameter~\citep{Sari:2000zp}, $\gamma_{\min}$ is given by Eq.~\ref{eq:gamma_min}, while $\gamma_{\rm{cool}}$ is given by dividing Eq.~\ref{eq:gamma_cool} by $1+Y$~\citep{Sari:2000zp}. We are using the notation $X_{y}= X/10^y$. Therefore, the cutoff $E_{\rm{KN}}$ varies over time and is usually larger at the onset of the afterglow.} 
This is a good approximation, since the VHE photons predominantly interact with low-energy protons, and the neutrinos produced in these interactions do not affect substantially the high-energy neutrino signal. 

\section{Hadronic interactions} \label{appedix1}
Because of the relatively small baryon density, $pp$ interactions are  subleading during the afterglow and only efficient  in the innermost regions of the outflow~\citep{Razzaque:2003uv, 2011MNRAS.415.2495M, Heinze:2020zqb}. Hence, the main channels for neutrino production are
\begin{eqnarray}
   p+ \gamma & \rightarrow & \Delta \rightarrow n + \pi^+, p + \pi^0  \\ 
p + \gamma & \rightarrow & K^+ + \Lambda/\Sigma \ .
    \label{reaction_channel}
\end{eqnarray} 
Neutral pions decay into gamma-rays $\pi^0 \longrightarrow 2 \gamma$, while neutrinos are produced through the charged pion decay $\pi^{+} \longrightarrow \nu_\mu + \mu^{+}$ followed by $\mu^{+} \longrightarrow \bar{\nu}_\mu + \nu_e + e^+$, and through $n \rightarrow p + e^- + \bar{\nu}_e$. Antineutrinos are also  produced  in the corresponding antiparticle channels; however, in this work, we do not distinguish between particle and antiparticles.

\subsection{Proton energy distribution}
Protons are assumed to be accelerated together with electrons at the forward shock driven by the blastwave in the cold CBM. Their comoving energy distribution is assumed to be [in units of GeV$^{-1}$ cm$^{-3}$]
\begin{equation}
n^\prime_p(E^\prime_p) = A^\prime_p E^{^\prime -k_p}_p \exp\biggl[-\biggl( \frac{E^\prime_p}{E^\prime_{p, \rm{max}}} \biggr)^{\alpha_p} \biggr] \Theta(E^\prime_p - E^\prime_{p, \rm{min}}) \ ,
\label{eq:proton_distribution}
\end{equation}
where $\Theta$ is the Heaviside function, $E^{\prime}_{p, \min}=\Gamma m_p c^2$~\citep{Dermer:2000yd, Murase:2007yt, Razzaque:2013dsa} is the minimum energy of accelerated protons and $E^\prime_{\max}$ is the maximum energy at which protons can be accelerated. The latter is fixed by equating the acceleration time scale of protons with their total cooling time, which takes into account all the energy loss mechanisms for accelerated protons. We refer the interested reader to Sec.~4 of~\cite{Guarini:2021gwh} for a detailed discussion.

Finally, $A^\prime_p = \varepsilon_p \xi_p u^\prime [ \int_{E^{\prime}_{p, \min}}^{E^{\prime}_{p, \rm{max}}} dE^\prime_p E^{\prime}_p n^{\prime}_{p}(E^\prime_p)]^{-1}$ is the normalization constant. Here, $u^\prime$ is the blastwave energy density defined in Eq.~\ref{eq:en_density}, $\varepsilon_p \lesssim 1 - \varepsilon_e - \varepsilon_B$ is the fraction of this energy which is stored into accelerated protons and $\xi_p$ is the fraction of accelerated protons.

The proton spectral index $k_p$ depends on the model invoked for particle acceleration. It is expected to be $k_p \simeq 2$~\citep{Matthews:2020lig} in the non-relativistic shock diffusive acceleration theory, while $k_p \simeq 2.2$ is expected from Monte-Carlo simulations of ultra-relativistic shocks~\citep{Sironi:2013ri}. The constant $\alpha_p=2$ mimics the exponential cutoff in the photon energy distribution~\citep{Hummer:2010vx}.

\section{Additional constraints on the properties of the circumburst medium}\label{sec:comparison}
Both the range of  $\tilde{E}_{k, \rm{iso}}$ allowed by the arguments in  Sec.~\ref{sec:compactness} and the  CBM density could span several orders of magnitude. 
A priori, it is not obvious whether our sample of VHE bursts (despite being based on a small number of bursts) shares common properties in terms of CMB densities with other GRBs without observed VHE emission. 

\cite{Gompertz:2018anr} performed a  scan of the parameter space allowed for  the blastwave isotropic energy and the CBM density for a selected set of GRBs not detected in the VHE regime. We stress that, in this appendix, we assume that  our sample of VHE GRBs (Table~\ref{tab:sample}) can be modelled by relying on the same assumptions as in \cite{Gompertz:2018anr} for the microphysical parameters. We also include GRB~130427A observed at $z=0.34$, with $\tilde{E}_{\gamma, \rm{iso}} \simeq 8 \times 10^{53}$~erg~\citep{2013GCN.14487....1G, Zhu:2013ufa}. Even though this burst has not been detected in the TeV range, it has been observed by \textit{Fermi}-LAT during the afterglow phase, with photons up to $\mathcal{O}(10)$~GeV about $9$~hours after the trigger~\citep{Zhu:2013ufa}. Being among the  most investigated events of this class, we consider GRB~130427A as representative of the HE sample observed  by  \textit{Fermi}-LAT~\citep{Ajello:2019zki}.

In light of the existing uncertainties on the microphysical parameters and in order to enable a comparison with the standard bursts of~\cite{Gompertz:2018anr} and the VHE ones considered in this work, we relax the values of the microphysical parameters considered in the main text and in Fig.~\ref{fig:constraints}. Our goal is to assess whether particular properties are preferred by GRBs emitting VHE photons with respect to standard GRBs.

Once the CBM type is fixed (ISM or wind), following~\cite{Gompertz:2018anr}, we  focus at  $11$~hours (as measured on Earth) after the trigger of the burst. At this time, two scenarios are possible: either $\nu_R< \nu_{\gamma, \rm{cool}} < \nu_X$ or $\nu_X < \nu_{\gamma, \rm{cool}}$, where  $\nu_R$ and $\nu_X$ are the observed effective  frequencies in the optical $R$ and X-ray bands, respectively. In the former case, we can infer the properties of the blastwave responsible for the afterglow emission~\citep{Sari:1997qe, Peer:2004doy, Gompertz:2018anr}: 
\begin{equation}
    \frac{\Phi^{\rm{obs}}_{R}}{\Phi^{\rm{obs}}_{{X}}} = \left( \frac{\nu_R}{\nu_X} \right)^{-k_e/2} \nu_R^{1/2} \nu_{\gamma, \rm{cool}}^{-1/2} \ , 
    \label{eq:inXray}
\end{equation}
where $\Phi^{\rm{obs}}_{R}$ and $\Phi^{\rm{obs}}_{X}$ are the  fluxes observed at $11$~hours in the $R$ and X-ray bands, respectively [both  in units erg~cm$^{-2}$~s$^{-1}$]. By replacing $\nu_{\gamma, \rm{cool}}$ in Eq.~\ref{eq:inXray} with Eq.~\ref{eq:synch_energies}, we obtain  a relation between  $\tilde{E}_{k, \rm{iso}}$ and $n_0$ or $A/(3 \times 10^{35})$.

If $\nu_X < \nu_{\gamma, \rm{cool}}$, the blastwave parameters can be inferred from the flux observed in the $R$ band. Plugging  $\Phi^{\rm{obs}}_{\nu, R}$ in the left hand side of Eq.~\ref{eq:photonFlux} and evaluating the right hand side of Eq.~\ref{eq:photonFlux} at $\nu_\gamma \equiv \nu_{R}$ provides us with a relation between $\tilde{E}_{k, \rm{iso}}$ and $n_0 \; [A/(3 \times 10^{35})]$~\citep{Sari:1997qe, Peer:2004doy}; see also Eqs.~$6$--$7$ in~\cite{Gompertz:2018anr}.

The  flux in the $R$ band is obtained by converting the AB magnitude through the following relation~\citep{1996AJ....111.1748F}:
\begin{equation}
    m_{\rm{AB}}= - 2.5 \log_{10} \left( \frac{\Phi^{\rm{obs}}_{\nu, R}}{3631} \right) \ ,
\end{equation}
where $\Phi^{\rm{obs}}_{\nu, R}$  is the observed flux [in units of  Jy]. The AB magnitudes are extracted from the~\cite{GCN}. 
The  flux at $11$~hours is extrapolated by evolving  $\Phi^{\rm{obs}}_{\nu, R} \propto t^{-\alpha_O}$, with $\alpha_O$ being the temporal spectral index in the optical band reported in Table~\ref{tab:betaO}. Note that these values do not include the intrinsic host galaxy extinction; hence, the value of $\Phi^{\rm{obs}}_{\nu, R}$ that we use is a lower limit of the real flux. We warn the reader that the value of $\alpha_O$ obtained for GRB 190114C from the standard closure relations and reported in Table~\ref{tab:betaO} does not  reproduce  the optical lightcurve and the spectral energy distribution simultaneously and satisfactorily. This hints that the standard afterglow model may not be adequate to model this GRB. Time-varying microphysical parameters might be more appropriate for this burst~\citep{Misra:2019vdg}; in this case our results would no longer hold.
 \begin{table}
    \centering
    \caption{Temporal optical index $\alpha_O$ obtained for the three considered VHE bursts and GRB 130427A, with relative References.}
    \begin{tabular}{ccc}
         Burst & $\alpha_O$ & References  \\ \hline
         GRB 180720B & 1.2 & ~\cite{Fraija:2019whb}\\
         GRB 190114C & 0.76 &~\cite{Fraija:2019wel} \\
         GRB 221009A & 0.52 &~\cite{2022GCN.32645....1B} \\ \hline
         GRB 130427A & 1.36 &~\citep{Panaitescu:2013pga} 
    \end{tabular}
    \label{tab:betaO}
\end{table}

We fix  the electron spectral index ($k_e$) as indicated in Table~\ref{tab:sample}. In particular, for GRB 190114C we fix the value $k_e=2.2$, while we checked that the results are not very sensitive to the variation of $k_e$. Furthermore,  we assume that the isotropic energy left in the blastwave after the prompt emission is $\tilde{E}_{k, \rm{iso}} \equiv \tilde{E}_{\gamma, \rm{iso}}$~\citep{Gompertz:2018anr}.  This implies a prompt efficiency of $\approx 50 \%$, which might be  optimistic~\citep{Beniamini:2016hzc} and  should  be rather interpreted as a lower limit on $\tilde{E}_{k, \rm{iso}}$.

In both the aforementioned regimes,  the microphysical parameters $\varepsilon_e$ and $\varepsilon_B$ should be fixed. \cite{Gompertz:2018anr} assumes $\varepsilon_e=0.1$ and $\varepsilon_B=10^{-1}$--$10^{-4}$ for all GRBs in their sample, and they conclude that $\varepsilon_B \simeq 10^{-4}$ is preferred  to avoid unphysical values of the CBM density.
The  parameters $\varepsilon_e=0.1$ and $\varepsilon_B = 10^{-5}$--$10^{-4}$ are also consistent with the typical values required for modelling the VHE emission through the SSC mechanism~\citep[e.g.][]{Fraija:2022rtj}.
We first rely on the same choice of the microphysical parameters of \cite{Gompertz:2018anr}  to favor a direct comparison between the properties of the VHE bursts and the standard ones and, to this purpose, we use $\varepsilon_e= 0.1$ and $\varepsilon_B=10^{-2}$--$10^{-4}$. Then, 
we  assume $\varepsilon_B = 10^{-5}$, while keeping $\varepsilon_e = 0.1$, since this value is allowed in the context of the SSC model. This procedure allows us to obtain  upper and lower limits for the CBM densities for the two underlying mechanisms. 

\begin{figure}
\centering
\includegraphics[width=0.45\textwidth]{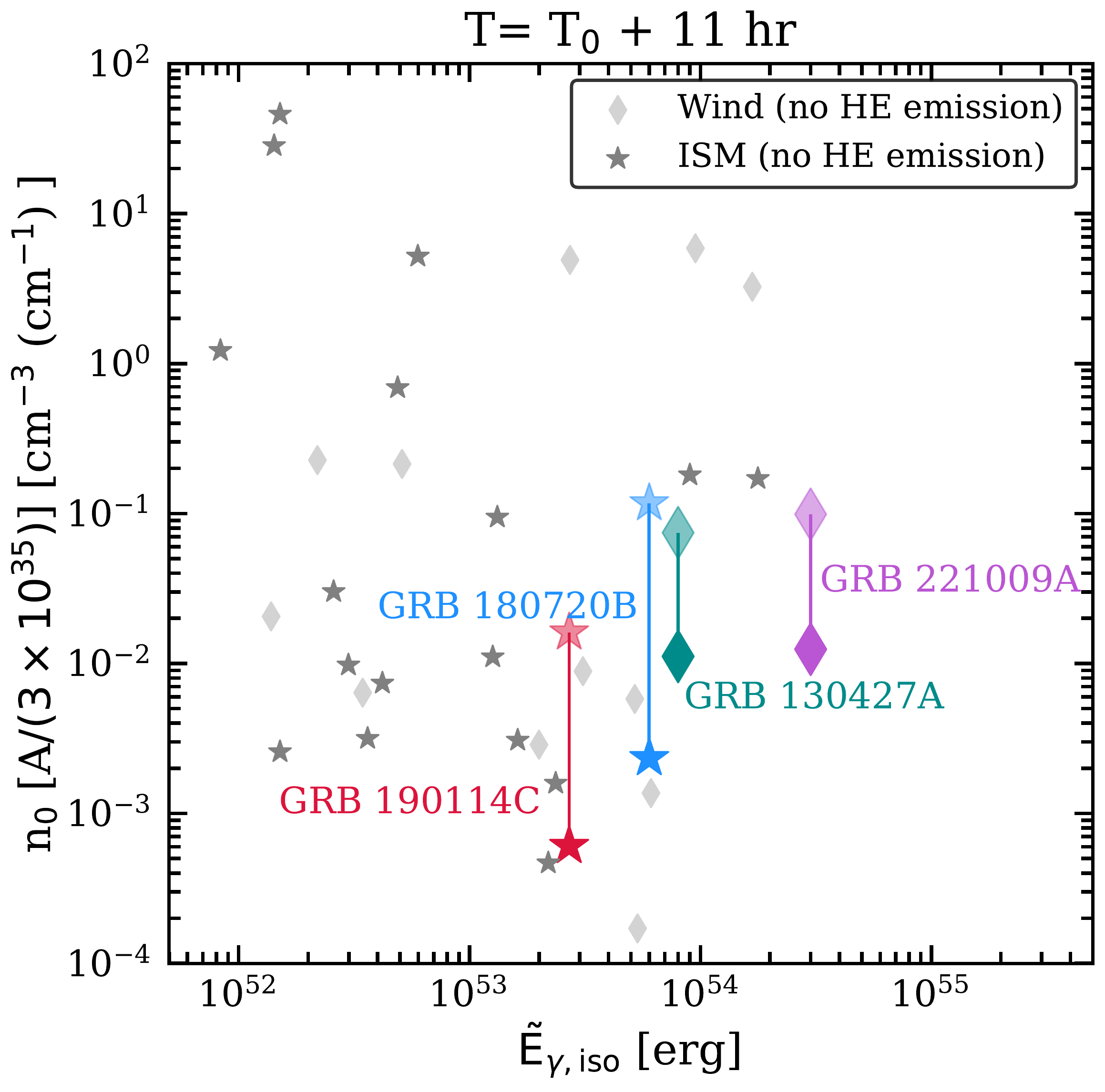}
\caption{VHE GRBs  in Table~\ref{tab:sample} (each distinguished by a different color) in the plane spanned by the isotropic gamma-ray energy $\tilde{E}_{\gamma, \rm{iso}}$
and CBM density [$n_0 \; \rm{or} \; A/(3 \times 10^{35})$]. 
Following~\protect\cite{Gompertz:2018anr}, we fix $\tilde{E}_{k, \rm{iso}} = \tilde{E}_{\gamma, \rm{iso}}$, $\varepsilon_e=0.1$ and $\varepsilon_B = 10^{-4}$--$10^{-5}$ for the lower (opaque markers) and upper limits (shadowed markers), respectively.
The stars denote bursts occurring in an ISM, while the diamonds correspond to bursts occurring in a wind-like CBM. The gray stars and diamonds correspond to the bursts  analyzed in~\protect\cite{Gompertz:2018anr} occurring in an ISM and a wind-like CBM, respectively, for $\varepsilon_e=10^{-1}$ and  $\varepsilon_B=10^{-4}$. The VHE GRBs in our sample favor low density CBM, for  $\varepsilon_B$ compatible with the SSC model; see main text for details.}
\label{fig:comparison}
\end{figure}
Figure~\ref{fig:comparison} summarizes our findings  for $\varepsilon_B= 10^{-5}$--$10^{-4}$. We include GRB 130427A in the plot, as representative of the GRBs detected in the HE regime during the afterglow; see Sec.~\ref{sec:events}.
The results obtained by adopting $\varepsilon_B = 10^{-4}$ can be directly compared to the ones of~\cite{Gompertz:2018anr}, as shown in Fig.~\ref{fig:comparison}  (gray markers). Intriguingly, the bursts detected in the VHE regime  cluster in the region of the parameter space corresponding to large isotropic energy emitted in gamma-rays and  relatively small CBM densities [$ 10^{-3}  \lesssim n_0 \lesssim 10^{-1}$ and $10^{-5} \lesssim A/(3 \times 10^{35}) \lesssim 10^{-1}$], consistently with our findings displayed in Fig.~\ref{fig:constraints}. 

The case with $\varepsilon_B=10^{-5}$ cannot be compared with the results in~\cite{Gompertz:2018anr} directly. Nevertheless, we consider it as representative of the  SSC model~\citep{Fraija:2022rtj}: since $n_0 \propto \varepsilon_B^{-(k_e+1)/2} \; [ A \propto   \varepsilon_B^{-(k_e+1)/4}]$, we expect the density to increase as $\varepsilon_B$ decreases, while keeping fixed $\tilde{E}_{k, \rm{iso}}$. It is worth noticing that decreasing $\varepsilon_e$  implies an  increase in the CBM density, because  $n_0 \propto \varepsilon_e^{(1-k_e)/2 } \; [A \propto \varepsilon_e^{(1-k_e)}]$, for the ISM [wind] scenario. For example, for $\varepsilon_e=10^{-2}$ and $\varepsilon_B=10^{-2}$, one obtains  results similar to the lower limits in Fig.~\ref{fig:constraints}. On the contrary, assuming $\varepsilon_e=10^{-2}$ and $\varepsilon_B = 10^{-4}$, shifts the points in Fig.~\ref{fig:comparison} to  larger densities, i.e.~$n_0 [A/(3 \times 10^{35})] \gtrsim 1$. Nevertheless, the multi-wavelength fits  in the literature suggest  $\varepsilon_e \simeq 0.1$. Hence, the densities obtained in Fig.~\ref{fig:comparison} might be preferred. 

We stress that the results in Fig.~\ref{fig:comparison} cannot be directly compared to the ones in Fig.~\ref{fig:constraints}, since in the former we fix $\tilde{E}_{k, \rm{iso}} = \tilde{E}_{\gamma, \rm{iso}}$, while in the latter $\tilde{E}_{k, \rm{iso}}$ is a free parameter. In Fig.~\ref{fig:constraints} the scaling of the CBM density with $\varepsilon_e$ and $\varepsilon_B$ is not trivial, since the isotropic kinetic energy is also changing with the other model parameters.

Note that,   $10^{-3} \lesssim \varepsilon_B \lesssim 10^{-1}$  (with $\varepsilon_e = 10^{-1}$) leads to   $10^{-8} \lesssim n \; [A/(3 \times 10^{35})] \lesssim 10^{-6}$, which is too low to be realistic~\citep{Gompertz:2018anr}. This result might be biased by theoretical limitations of the closure relations and by the assumption $\xi_e=1$. 
While the arguments in Sec.~\ref{sec:vhe} are not constraining for GRB 180720B,  we conclude from Fig.~\ref{fig:comparison} that low densities might be preferred for VHE bursts for typical microphysical parameters consistent with a SSC scenario, as also found in~\cite{Wang:2019zbs}.


\bsp	
\label{lastpage}
\end{document}